\documentclass[iop]{emulateapj-rtx4} 
\usepackage{natbib,amsmath,lscape}

\usepackage{lscape}
\usepackage{graphicx}
\usepackage{natbib}
\usepackage{amssymb}
\usepackage{float}

\usepackage{makeidx}
\usepackage{amsmath}
\usepackage{rotating}
\usepackage{longtable}
\usepackage{subfigure}
\usepackage{subfloat}
\usepackage{morefloats}
\usepackage{captcont}
\nocite{*}

\newcommand{\etal}{et~al.\/}

\newcommand{\lya}{\mbox{Ly$\alpha$}}
\newcommand{\cm}{cm$^{-2}$}

\newcommand{\ovi}{\hbox{O\,{\sc vi}}}
\newcommand{\nii}{\hbox{N\,{\sc ii}}}
\newcommand{\niii}{\hbox{N\,{\sc iii}}}
\newcommand{\sii}{\hbox{Si\,{\sc ii}}}

\newcommand{\siii}{\hbox{Si\,{\sc iii}}}
\newcommand{\siiv}{\hbox{Si\,{\sc iv}}}
\newcommand{\nv}{\hbox{N\,{\sc v}}}
\newcommand{\siovi}{\siiv/\ovi}
\newcommand{\nvovi}{\nv/\ovi}
\newcommand{\novi}{\rm N_{\rm O\,{\sc VI} }}
\newcommand{\nnv}{\rm N_{\rm N\,{\sc V} }}
\newcommand{\nsiiv}{\rm N_{\rm Si\,{\sc IV} }}
\newcommand\mlstar{L^*}
\newcommand\lstar{$\mlstar$}

\def\cm#1{\, {\rm cm^{#1}}}

\shortauthors{Werk \etal}
\shorttitle{Origins of the Highly-Ionized CGM}

\begin{document} 
\slugcomment{v2.0}

\title{The COS-Halos Survey: Origins of the Highly Ionized Circumgalactic Medium of Star-Forming Galaxies}

\author{Jessica K.\ Werk\altaffilmark{1,2},
  J. Xavier Prochaska\altaffilmark{2},
 Sebastiano Cantalupo\altaffilmark{3, 2},
  Andrew J. Fox\altaffilmark{4},
  Benjamin Oppenheimer\altaffilmark{5},
  Jason Tumlinson \altaffilmark{4}, 
  Todd M. Tripp\altaffilmark{6},
  Nicolas Lehner\altaffilmark{7}, \&
  Matthew McQuinn\altaffilmark{1}
  }

\altaffiltext{1}{University of Washington, Department of Astronomy, Seattle, WA $jwerk@uw.edu$ }
\altaffiltext{2}{UCO/Lick Observatory; University of California, Santa Cruz, CA }
\altaffiltext{3}{Institute for Astronomy, ETH Zurich, Wolfgang-Pauli-Strasse 27, 8093 Zurich, Switzerland}
\altaffiltext{4}{Space Telescope Science Institute, 3700 San Martin Drive,  Baltimore, MD}
\altaffiltext{5}{CASA, Department of Astrophysical and Planetary Sciences, University of Colorado, Boulder, CO 80309}
\altaffiltext{6}{Department of Astronomy, University of Massachusetts, Amherst, MA}
\altaffiltext{7}{Department of Physics, University of Notre Dame, Notre Dame, IN 46556}

\begin{abstract}

The total contribution of diffuse halo gas to the galaxy baryon budget strongly depends on its dominant ionization state. In this paper, we address the physical conditions in the highly-ionized circumgalactic medium (CGM) traced by  \ovi\ absorption lines observed in COS-Halos spectra. We analyze the observed ionic column densities, absorption-line widths and relative velocities, along with the ratios of  \nvovi\ for 39 fitted Voigt profile components of \ion{O}{6}. We compare these quantities with the predictions given by a wide range of ionization models. Photoionization models that include only extragalactic UV background radiation are ruled out; conservatively,  the upper limits to \nvovi\ and measurements of  N$_{\rm OVI}$ imply unphysically large path lengths $\gtrsim$ 100 kpc.  Furthermore, very broad \ovi\ absorption (b $>$ 40 km s$^{-1}$) is a defining characteristic of the CGM of star-forming L$^{*}$ galaxies. We highlight two possible origins for the bulk of the observed \ovi: (1) highly structured gas clouds photoionized primarily by local high-energy sources or (2) gas radiatively cooling on large scales behind a supersonic wind. Approximately 20\% of circumgalactic \ion{O}{6} does not align with any low-ionization state gas within $\pm$50 km s$^{-1}$ and is found only in halos with M$_{\rm halo}$ $<$ 10$^{12}$ M$_{\odot}$. We suggest that this type of unmatched \ion{O}{6} absorption traces the hot corona itself at a characteristic temperature of 10$^{5.5}$ K.  We discuss the implications of these very distinct physical origins for the dynamical state,  gas cooling rates, and total baryonic content of L$^*$ gaseous halos.

\end{abstract}


\section{Introduction}
\label{sec:intro}

Quasar absorption-line techniques have established that present-day galaxies
are enveloped by a highly-ionized and enriched plasma extending to hundreds of kpc.
Although largely invisible in its faint emission, this `halo gas' or `circumgalactic medium' (CGM) 
is revealed by observations of the \ovi\ doublet at $\lambda\lambda
1031,1037$ in far-UV absorption-line spectroscopy of background quasars  
\citep[e.g.][]{tripp08, wakker09,pwc+11,tumlinson11, savage14}.
Dedicated studies have assessed 
the covering fraction, surface density, and radial extent of this
\ovi-bearing gas, including its relationship to galaxy  
stellar mass and star formation rate. While this \ovi\ is common 
around $\sim L^*$ star-forming galaxies, it appears less 
common around non-star-forming galaxies \citep{tumlinson11} and  
low-mass dwarf galaxies \citep{pwc+11}. It may \citep{tripp00, shull03, chen09, stocke14, johnson15} or may not 
\citep{wakker09} be common in galaxy-group environments and/or intracluster gas \citep[e.g.][]{bowen01}.

Recent observational studies have placed lower limits on the  highly-ionized  metal mass in the CGM of  L$^{*}$ galaxies traced by \ovi. It most likely exceeds 10$^{7}$ M$_{\odot}$ \citep{tumlinson11}, comparable to or greater than the metal mass within their ISM \citep{peeples14}.  The chief uncertainty in CGM metal-mass estimates arises from the largely unknown ionization conditions of the gas. Typically, lower limits on oxygen budgets in the highly ionized gas are derived by  simply assuming that \ovi\ is at its maximum ionization fraction in typical low-density conditions. However,  the true metal budgets vary greatly depending on the model used to explain the ionization, which in turn significantly affects conclusions about the fate and origin of the highly ionized gas around galaxies.

Several authors have recently addressed the origin and total baryonic content of the highly ionized gas in the CGM by comparing simulations with observations.  Strong feedback processes (from both star-formation and AGN) give rise to significant  \ovi\ absorption in the CGM of both cosmological zoom-in and hydrodynamical simulations \citep[e.g.][]{shen13, hummels13, cen13, suresh15, liang16, ford16, oppenheimer16, rahmati16, cen16}. However, the column density of the simulated \ovi\ absorption depends upon an assumed ionization mechanism of the gas. Just as observers must model their absorption-line data with a radiative transfer code like Cloudy (Ferland et al. 2013) to determine the physical characteristics of the gas (e.g. Werk et al. 2014), so must simulators reproduce the physical conditions in their gas cells using radiative transfer models that provide corresponding values of gas column density of a given ion. Under the standard assumption of ionization equilibrium and a combination of photo- and collisional ionization, simulators compare their model-derived radial distributions of  \ovi\ absorption with the \ovi\ column density distributions from observations (though see Oppenheimer et al. 2016). These comparisons routinely reveal a deficit of \ovi\ column density in the simulated gas cells compared to observations, perhaps hinting at some unaccounted for ionization process \citep[e.g.][]{suresh15}.

It has thus far been difficult to generalize the physical conditions of intergalactic and/or circumgalactic gas bearing \ovi\ despite its frequent detection and broad characterization in UV spectroscopic observations.  Detailed studies have been carried out on an absorber-by-absorber basis in high-resolution and high-S/N QSO spectra that trace low-density foreground gas in the IGM and galaxy halos \citep{richter04, sembach04,  tripp06, tripp08, howk09, narayanan10, narayanan11, savage11, narayanan12, savage14, muzahid15, pachat16}. These studies have revealed the ionization mechanisms of \ovi\ and other `high-ions' like \nv\ to be both varied and complex over a wide range of environments \citep[e.g.][]{sembach04}.  Line diagnostics from low, intermediate, and high ions, including ionic column density ratios and absorption-line profiles, sometimes support  a similar, photoionized origin for \ovi, \nv,  and low-ionization state gas \citep[e.g.][]{tripp08, muzahid15}, and sometimes require \ovi\ to be ionized by collisions of electrons with ions in a $\sim$10$^{5.5}$ K plasma \citep[e.g.][]{tumlinson05, fox09, savage11,tripp11, wakker12,  narayanan12, meiring13}. Often, the multiple components for a single absorber show both narrow and broad absorption lines consistent with both scenarios.  A similar challenge arises from the diffuse gas known as high velocity clouds (HVCs) in the halo of the Milky Way \citep{sembach03, fox04,fox05,fox06, lehner09, wakker12}.    In this case, the complex mixing and shocking of cooler gas within a hot, ambient medium is often consistent with the absorption-line diagnostics.

\begin{figure*}[t!]
\begin{centering}
\hspace{0.1in}
\includegraphics[width=0.91\linewidth]{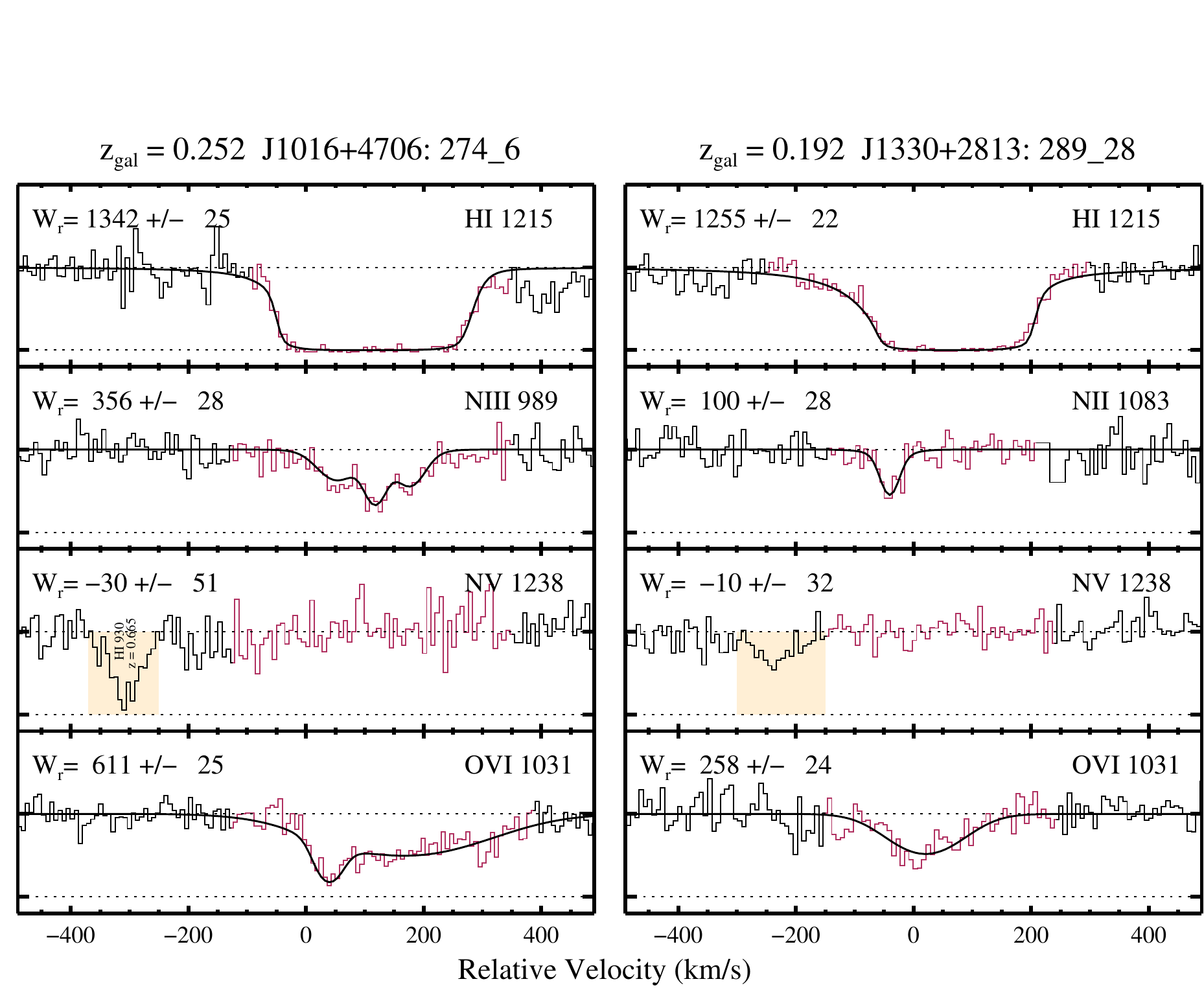}
\end{centering}
\caption{Ionic species stacks drawn from the {\emph{HST}}/COS absorption-line spectra, centered on the transitions  relevant to this study: HI Ly$\alpha$, \ion{N}{2} or \ion{N}{3}, NV, and \ion{O}{6}. We show data for two representative galaxies, J1016$+$4706: 274\_6 at $z=0.252$, and J1330$+$2813: 289\_28 at $z = 0.192$. In each panel, v $=$ 0 corresponds to the transition wavelength at the  galaxy systemic redshift. Generally,  the associated CGM absorption falls within $\pm 300$ km s$^{-1}$ of the galaxy systemic redshift. In both of these examples, we see strong absorption from HI, the low ionization state transitions of nitrogen, and OVI, but NV is undetected. The light shaded area on each NV panel marks a feature that is not due to NV in either case, but absorption from a system at some other redshift. For example, in the NV panel associated with J1016$+$4706: 274\_6, this absorption line is part of the Lyman Series (\ion{H}{1} $\lambda$= 930 \AA) for an absorption-line system at z $=$ 0.665.}
\label{fig:exspec}
\end{figure*}

\begin{figure*}[t!]
\begin{centering}
\includegraphics[width=0.75\linewidth, angle =90.0]{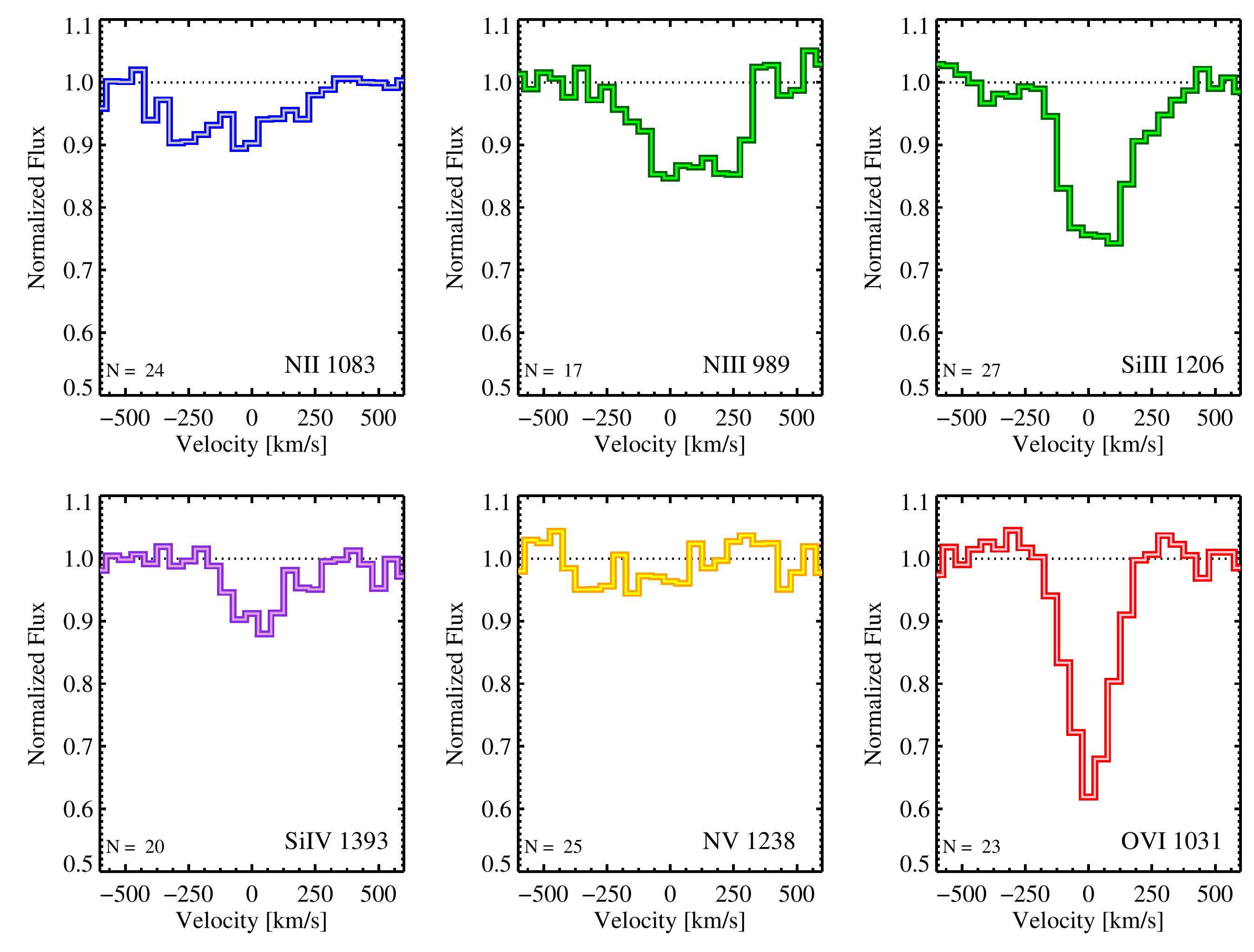}
\end{centering}
\caption{Average COS-Halos absorption-line spectra for NII, NIII, SiIII, SiIV, NV, and OVI,  stacked at v $=$ 0 km s$^{-1}$ corresponding to the galaxy systemic redshift. Only the COS-Halos star-forming galaxies are included in these stacks.  The resultant stacks may include coincident absorption from unrelated gas at different redshifts (or, in the case of \ion{N}{3}~989, a neighboring transition from \ion{Si}{2}), but this is a minor effect. The number of spectra averaged in each panel is shown in the lower-left corners. We use a spectrum if there is coverage of the transition. We note that strong absorption features (SiIII, OVI) are not driven by a small set of events, since a median stack yields qualitatively similar results.
}
\label{fig:stacks}
\end{figure*}

 
Here, we focus on the physical conditions of the \ion{O}{6}-bearing gas around z$\sim$0.2 star-forming $L \approx \mlstar$ galaxies. Previous studies have focused on the Milky Way itself \citep[e.g.][]{fox04, lehner11, wakker12}, on single sightlines with exquisite, high-S/N UV spectra \citep[e.g.][]{narayanan10, narayanan11, tripp11, narayanan12, meiring13}, z$\sim$ 2$-$3 Lyman limit systems or damped Lyman $\alpha$ systems \citep{fox09, lehner14}, or on absorption lines originating in wide variety of galaxy or group environments \citep{heckman02, grimes09,  wakker09, bordoloi16}. The COS-Halos dataset provides a uniform sample of absorbers with well-characterized host galaxy properties that allow us to generally constrain ionization processes affecting \ovi\ in the star-forming galaxy halo environment and potentially relate them to galaxy properties.  While the COS-Halos spectra have only moderate S/N ($\sim$10 at \ovi ), they cover a wide range of transitions at z$\sim$ 0.2, including (but not limited to) \sii, \siii, \siiv,  \nii, \niii, \nv, and \ovi. The coverage of such a variety of ionic species allows us to uniquely assess the multiphase nature of the gas within 150 kpc of a star-forming galaxy for 24 distinct sightlines. Ultimately, our goal is to apply the best available diagnostics from the COS-Halos dataset \citep{tumlinson13, werk13} to the origins of the highly ionized gas in gaseous galactic halos, and from these diagnostics to draw conclusions about the gas flows driving evolution in these galaxies.


In Section 2 we describe the data and review the relevant properties of the low-ionization state gas. In Section 3, we present a detailed Voigt profile-based kinematic analysis, joint with the low-ions,  of the individual  \ion{O}{6} absorbers, and examine several significant correlations between the gas kinematics and COS-Halos host galaxy properties. Section 4 presents an analysis of the many ionization processes capable of producing a highly ionized plasma and compares their predictions with the COS-Halos data for component column densities, gas velocities,  line-widths, and ratios of \siovi\ and \nvovi.   In Section 5, we summarize our results. Finally, in Section 6 we address several recent analytical and phenomenological models and comment on their applicability to our findings and implications for the co-evolution of the galaxy and its CGM. 

\section{Observed Gas Properties}
\label{sec:data}

\subsection{Data and Sample}

The observations of CGM gas are taken from the COS-Halos survey,
a dedicated absorption-line survey of $L \approx \mlstar$ galaxies
at $z \sim 0.2$ \citep{werk13, tumlinson13}. 
The survey targeted 44 galaxies with a diversity of
SFRs and morphology \citep{werk12}, using background quasars at
impact parameters $R \approx 10-160$\,kpc.  
Far-UV spectra were obtained with the Cosmic Origins Spectrograph
\citep[COS;][]{froning09, green12} on the {\it Hubble Space Telescope} (PID 11598; PI Tumlinson). The data cover $\lambda_{\rm obs}
\approx 1150-1600$\AA, and are complemented by ground-based,
echelle spectroscopy (Keck/HIRES) of the background quasar, long-slit spectroscopy
of the targeted galaxies (and others in the field), and 
SDSS photometry \citep{werk12}.  

Because the COS optics do not correct for the mid-frequency
wave front errors arising from zonal irregularities in the HST
primary, the true COS line-spread function (LSF) is not characterized
by a single Gaussian. Instead, it is well described by
a Gaussian convolved with a power law that extends to many
tens of pixels beyond the line center \citep{coslsf}. 
These broad wings affect both the precision of our equivalent
width measurements and complicate assessments of line saturation.
We mediate these effects when we fit absorption lines
(described in Section 3) by using the nearest wavelength grid
point and convolving with the real LSF. Each
COS resolution element at R  $\sim$18,000 covers 16  km s$^{-1}$ and
is sampled by six raw pixels.  We performed our analysis on the data binned by
three native spectral pixels to a dispersion of $\Delta\lambda$ $\approx$ 0.0367 \AA. The resulting science-grade spectra are characterized by  S/N $\sim$ 10$-$12 per COS resolution element which has FWHM $\approx$ 18 km s$^{-1}$. We assume an absolute velocity calibration error due to the systematic uncertainty in the COS wavelength calibration of $\pm$10 km s$^{-1}$ \citep{tumlinson13}.

In this paper, we focus on the 24 star-forming galaxies that exhibit positive detections of \ion{O}{6} in their inner-CGM.\footnote{Officially, there is a 25th COS-Halos galaxy that would meet the criteria for star-forming host galaxy and detected \ovi, J1437$+$5045 317\_38. However, the COS spectrum of this QSO has an anomalously low S/N (S/N $\approx$ 4) compared to the rest of our dataset, and we therefore exclude it from our analysis.}  We omit from our study the 4/13 non-star-forming galaxies that show positive detections of \ion{O}{6}\footnote{At least two of these systems have less-massive star-forming galaxies within 200 kpc that complicate the interpretation (Werk et al., in prep)} primarily because we wish to carry out a controlled study unaffected by environmental effects; massive non-star-forming galaxies are known to live in high density environments.  We additionally note that there are 2 star-forming galaxies in the COS-Halos sample that show no \ovi\ absorption in their halos. Each is an interesting system in its own right. The first,  J1550$+$4001 97\_33, has an impact parameter of 151 kpc and shows only very broad HI Ly$\alpha$ absorption (Doppler $b$-parameter $\approx$ 90 km s$^{-1}$) with log N$_{\rm HI}$ = 13.95 cm$^{-2}$. The second, J0943$+$0531 106\_34, lies at an impact parameter of 120 kpc and has a  Lyman Series down to Ly$\gamma$ giving log N$_{\rm HI}$ = 15.5 cm$^{-2}$. The only detected metal ion in this sightline is \ion{Si}{3}, and the estimated metallicity is very close to solar (Prochaska et al., in prep). The 2$\sigma$ detection limit for \ion{O}{6} corresponds to a typical column density upper limit of 10$^{13.5}$ cm$^{-2}$. All of the components we consider in this analysis are well-detected ($>$ 3$\sigma$ detections). 

 Twenty-two of the 24 galaxies in our sample exhibit positive detections of at least one low-ionization state metal line (e.g. \ion{Si}{2}, \ion{Mg}{2}, \ion{C}{3}, \ion{Si}{3}) in addition to \ovi. In every case, the COS spectra also provide spectral coverage of the \ion{N}{5} doublet, and many have coverage of \ion{Si}{4}, both previously analyzed by \cite{werk13}. Figure~\ref{fig:exspec} shows two representative examples of the
complete dataset.   We note the strong detections of \ion{O}{6} and the non-detection of
\ion{N}{5} absorption despite the obvious presence of
\ion{N}{2} and \ion{N}{3}.  Of the 23 galaxies covering both 
\ion{O}{6} and  \ion{N}{5} in COS-Halos, only  3/23 have positive detections
of \ion{N}{5}. The typical 2$\sigma$ upper limits on log $\nnv$ range from 13.4 $-$ 13.8 cm$^{-2}$. Of the 19 star-forming galaxies whose COS spectra cover both \ion{O}{6} and \ion{Si}{4}, 8/19 exhibit positive detections of  \ion{Si}{4}. These detection rates contrast with the more frequent detection of Si and N through lower-ionization transitions of \ion{Si}{2}, \ion{Si}{3}, \ion{N}{2}, and \ion{N}{3}. Table 3 of Werk et al. 2013 summarizes the
constraints on the column densities for these three ions.\footnote{The  COS spectra cover the \ion{C}{4} doublet in a few cases, but the S/N  is too poor at its wavelength to offer a meaningful constraint.}

Figure~\ref{fig:stacks} illustrates the average absorption profiles of
the sample of COS-Halos sightlines considered here.   We have constructed
these ``stacks''
by averaging the spectra in the rest-frame of each
targeted star-forming galaxy \citep[see][for details]{qpq6}\footnote{
The resultant stacks include coincident absorption from unrelated gas
at different redshifts (or, in the case of \ion{N}{3}~989, a
neighboring transition), but this is a minor effect.}. 
The stacked profiles show strong detections of \ion{O}{6} absorption
and also \ion{Si}{3}, \ion{N}{2}, and \ion{N}{3}, consistent with
the results for individual systems.  On average, however, the data
exhibit no detectable absorption at \ion{N}{5} ($\lambda\lambda$1328,1242) and
weak absorption at the \ion{Si}{4} ($\lambda\lambda$1393, 1402) doublets, which have upper limits of $W_{\rm r} <  100, 80$m\AA, respectively).


\subsection{Abundance Ratios Relative to Solar Values}
\label{sec:nitrogen}

The non-detection of \ion{N}{5} for the vast majority of our sightlines may  indicate that the detected CGM gas is under-abundant in nitrogen, especially with respect to oxygen. The COS-Halos mean value for the {\emph{upper limit}} to  log ($\nnv / \novi$) is $-0.9$, and ranges from $0.0$ to $-1.9$. Ignoring ionization effects, this typical upper limit is comparable to the solar value for log(N/O) of $-0.86$ \citep{asplund09}.  We note that 17/39 of the COS-Halos absorbers with spectra covering both \ion{O}{6} and  \ion{N}{5} demand that  log ($\nnv / \novi$) $<$ $-0.9$.  As we will discuss in subsequent sections, different ionization conditions can strongly impact the conversion between log ($\nnv / \novi$)  and log(N/O). Some models will indicate that the COS-Halos upper limits to log ($\nnv / \novi$) require the assumption of anomalously low log(N/O) relative to the solar value to give physically plausible results. In contrast, other ionization models will imply that our non-detection of  \ion{N}{5} is consistent with the solar ratio of N/O for a large range of physically plausible conditions.  We may use our limits on log ($\nnv / \novi$) to constrain the ionization state of the gas {\emph{only}} under the assumption of solar relative abundances. Here, we use independent diagnostics to examine whether the assumption of solar relative abundances in the CGM is approximately valid.

 We point out that the N/O ratio is sometimes found to be sub-solar in metal-poor environments that include the ionized gas of \ion{H}{2} regions \citep{henry93} and the neutral gas damped \lya\ systems \citep[DLAs; e.g.][]{reimers92,pho+02, battisti12, zafar14},  sub-DLAs \citep[e.g.][]{tripp05, battisti12}, and Lyman Limit systems \citep[e.g.][]{jenkins05}. There are counter examples in which  \ion{N}{5}  has been detected in analogous highly ionized absorbers along with other ions that enable the relative nitrogen abundance to be checked (e.g. SV, NeVIII, CIV, OVI). In  these cases, there is no evidence of a nitrogen under-abundance \citep{savage02, tripp06, tripp+11}. 

Figures~\ref{fig:exspec} and \ref{fig:stacks}
demonstrate that nitrogen is routinely detected
in the COS-Halos sample, via both the \ion{N}{2}~1083 and \ion{N}{3}~989
transitions.  Furthermore, \cite{werk14} found that the observed
ratios of \nii~and \niii~column densities are consistent
(within the 0.3 dex uncertainty) with the solar relative abundance photoionization modeling that includes low ionization states of carbon, silicon, and magnesium.

\begin{figure}[h!]
\begin{centering}
\hspace{-0.3in}
\includegraphics[width=1.1\linewidth]{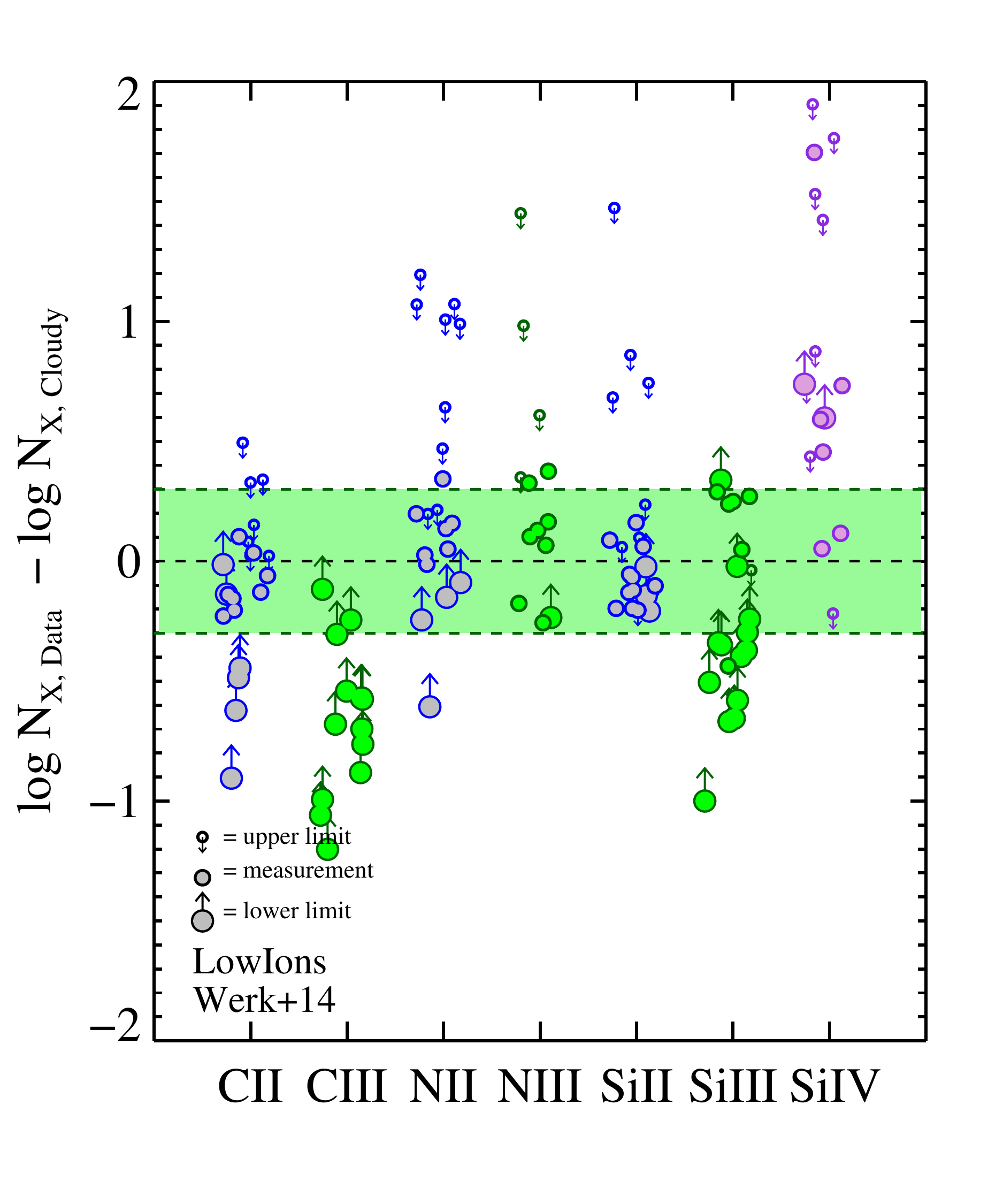}
\label{fig:photo}
\end{centering}
\vspace{-0.2in}
\caption{ The difference between the measured ionic column densities and the column densities from the best-fitting Cloudy models, described in detail in Werk et al. 2014. Lower limits (indicating saturation in the absorption-line profile), are marked as the large circles with upward arrows. Upper limits (i.e. non-detection of a transition) are marked by the open small circles with downward-facing arrows. In cases where we can measure unsaturated detections, we mark the points as medium-sized filled circles. Data are color coded for the specific ionization states of the metal lines, listed on the x-axis. We consider \ion{Si}{4} to be an intermediate-to-high ion, and as such, the Cloudy model solutions based on the low-ions tend to underproduce the typical \ion{Si}{4} column densities we see in the COS data. The variety of low-ionization state transitions are consistent with solar ratios in gas photoionized by an EUVB to within $\pm$0.3 dex. 
}
\label{fig:models}
\end{figure}

More directly,
Figure~\ref{fig:models} presents a comparison between the predicted
column densities of \nii~and \niii~for the systems using the
photoionization models described by \cite{werk14} in which the CGM is ionized primarily by the extragalactic UV background \citep[EUVB; ][]{hm01}.  The resultant
ionization parameters range from $ -2 < \log U < -4$, 
values which are common for lower-ionization state gas \citep[e.g.][]{p99,lehner13}. These models were
constructed to match the constraints from all of the lower ionization
states of the CGM, including the \ion{H}{1} gas.  

The y-axis of Figure \ref{fig:models} shows the difference between the measured ionic column densities and the column densities from the best-fitting Cloudy models presented in \cite{werk14}. We show the data minus model differences for the low-ions most useful in constraining the solution for the ionization state of the $T \approx 10 ^{4}$ K gas:  \ion{Si}{2}, \ion{Si}{3}, \ion{N}{2}, and \ion{N}{3}.  Absorption lines that are saturated in the COS spectra are shown as lower limits (upward-facing arrows), while the two sigma non detections are given as upper limits (downward-facing arrows). The models tend to simultaneously fit the many transitions of low-ion data well, with a characteristic systematic error of $\pm$0.3 dex. The underlying assumptions that characterize these Cloudy models are: (1)  a constant gas density, (2) photoionization and thermal equilibrium, and (3) plane-parallel geometry.  The systematic error is dominated by the uncertainty in the slope of the ionizing spectrum, and to a lesser extent, the departure of the elemental abundances from solar ratios. As discussed later in this paper and by \cite{stern16}, introducing a density gradient into the Cloudy models tends to bring the models into even better agreement with the data. In such a model, even adjacent ionization states of metal lines like  \ion{Si}{2} and \ion{Si}{3} arise in hierarchically-nested gas clouds characterized by different densities. We note that \cite{stern16} assume solar ratios of the elements, and quote an uncertainty of $\pm$0.1 dex that accommodates possible departures from solar ratios in the COS-Halos data with their model. 


Figure~\ref{fig:models} shows broadly that these single-density photoionization models are consistent with
the observed column densities of \nii~and \niii~under the
assumption of solar relative abundances. Thus, anomalously low Nitrogen abundances are not necessary to explain the column densities of the low ionization state gas within the CGM.   We proceed under the
expectation of approximately solar relative abundances for the more highly ionized plasma traced by \ion{O}{6}
and \ion{N}{5}.  To first order,  the model predictions scale directly with this assumption. 

\section{Gas Kinematics}
\label{sec:kinematics}

\begin{figure*}[t!]
\begin{centering}
\hspace{0.2in}
\includegraphics[width=0.90\linewidth]{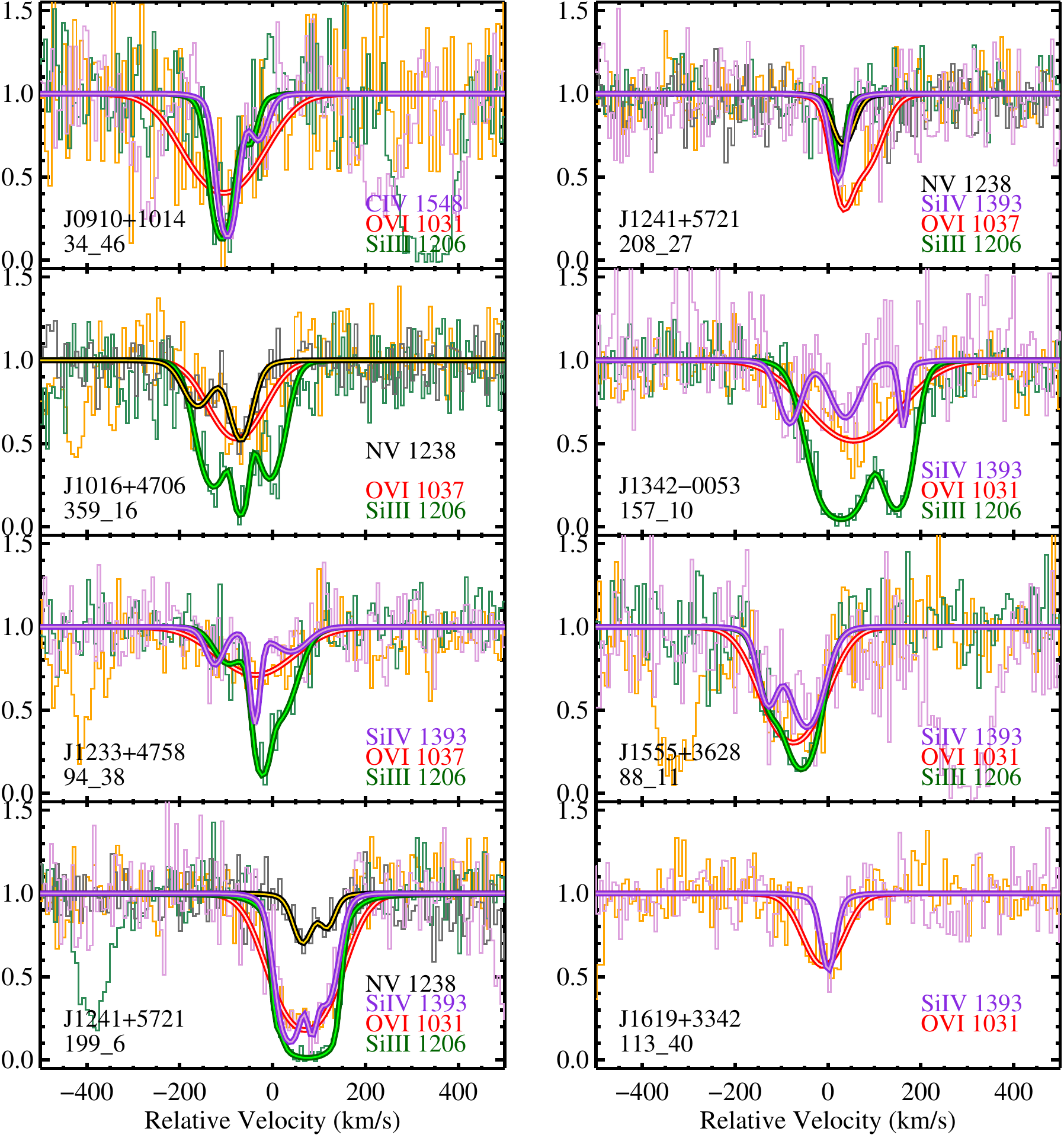}
\end{centering}
\caption{ Comparison of  continuum-normalized profile-fits for  \ion{O}{6} (red/orange), \ion{N}{5}, (black/yellow),  \ion{Si}{4} or \ion{C}{4} (purple), and \ion{Si}{3} (green) for 8 selected sightlines. For reference, data for each ion is shown in the same color as its Voigt profile fit.  A relative velocity of 0 km s$^{-1}$ corresponds to the systemic redshift of the associated host galaxy, determined by ground-based optical spectroscopy \citep{werk12}. In the lower-left corner, we provide the COS-Halos QSO name and the galaxy ID, which can be easily cross-referenced with spectra and other galaxy properties presented in previous COS-Halos papers. 
}
\label{fig:highprof}
\end{figure*}

\subsection{The Potential for Multiple Gas Phases}

Figure \ref{fig:models} additionally shows the difference between COS-Halos measurements and Cloudy column density predictions for the intermediate-ion \ion{Si}{4} using the same single-phase photoionization model as the one that matches the low-ions. This best-fitting low-ion Cloudy photoionization model under-predicts log $\nsiiv$ for 6/8 detected \ion{Si}{4} lines by 0.5 $-$ 1.8 dex.  \cite{werk14} note this obvious discrepancy between the  \ion{Si}{4} measurements and model predictions, and point out that the additional \ion{Si}{4} seen in the data can be explained by  invoking an intermediate-ionization gas phase. Furthermore, as one might expect, this issue is resolved by imposing a steep, inwardly-increasing density gradient on the absorbing gas cloud. The result is that \ion{Si}{4} represents lower-density, photoionized outer-layer gas \citep{stern16}. 

To an even greater extent, the same discrepancy exists for \ion{O}{6} if we model it with the low-ions (not shown on Figure \ref{fig:models}; it lies beyond the upper y-axis bound at $>$ 2 dex difference). Many previous studies have recognized this need for an additional highly-ionized gas phase  \citep{tripp+11, tumlinson11, meiring13, lehner13}, and it has become common parlance to refer to the CGM as `multiphase' for this reason (and others, outlined in the next section).   Whether or not \ion{Si}{4} is in the same gas phase as \ion{O}{6}, or \ion{O}{6} represents an even lower density outer-cloud layer,  is a question that we explore in later sections. Here,  we simply point out the apparent excess of \ion{Si}{4}, and to a greater extent, \ion{O}{6} with respect to the predictions of single-phase, single-density, photoionization equilibrium models that include ionizing radiation from only the EUVB. 


In addition to the column density ratios discussed in the previous section, several properties of CGM absorption line shapes lead to the common conclusion of a multiphase CGM.  One piece of kinematic evidence for multiple phases, summarized below, is the occasional misalignment of  the high and low-ion absorption profiles in velocity space (see also Fox et al. 2013 \nocite{fox13}). Another is the systematically broader line widths of \ovi~absorption \citep[e.g.][]{tumlinson05, lehner09, narayanan10, savage11, muzahid12, stocke13}. The two dominant gas phases are typically described as a cool $\sim$10$^{4}$ K phase, and a warm,  $\sim$10$^{5.5}$ K phase \citep{tripp08}. Though we are concerned primarily with the highly ionized gas in this analysis (traced by \ovi ), the models we present have bearing on its interplay with the cooler gas phase traced by low (and intermediate) ionization state absorption lines.  Thus, the gas kinematics of both the low and high-ions impose an important observational constraint on each of the models. 

%


\begin{figure*}[t!]
\vspace{0.2in}
\begin{centering}
\hspace{0.6in}
\includegraphics[width=0.85\linewidth]{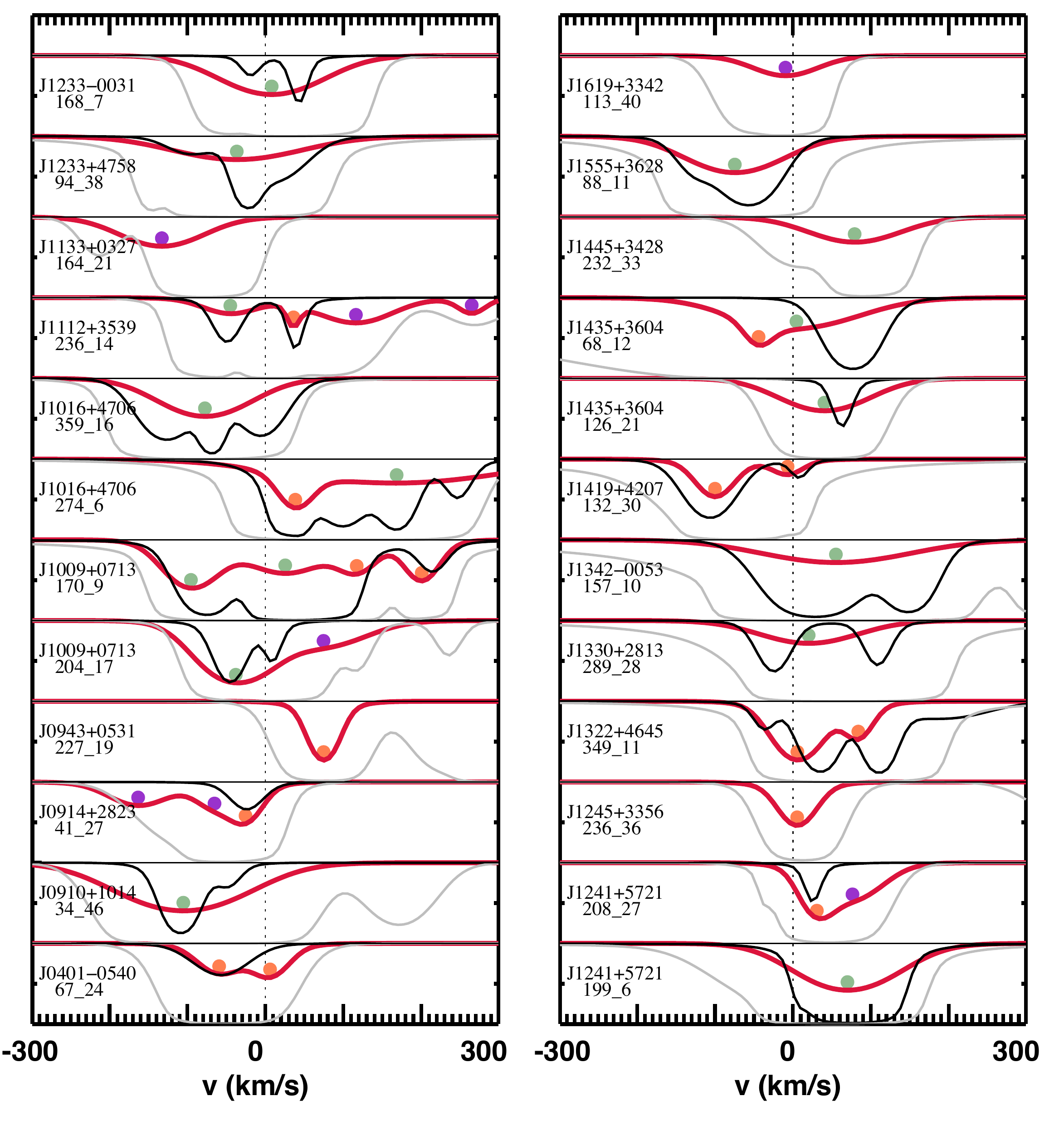}
\end{centering}
\caption{ Comparison of  continuum-normalized profile-fits for  \ion{O}{6} (dark red),  \ion{H}{1} (gray), and \ion{Si}{3} (black) for all 24 sightlines included in this work. On the x-axis is velocity relative to the systemic velocity of the associated host galaxy, determined by ground-based optical spectroscopy \citep{werk12}. To the left of the profiles, we provide the COS-Halos QSO name and the galaxy ID, which can be easily cross-referenced with spectra and other galaxy properties presented in previous COS-Halos papers. The filled, colored circles at the center of each \ion{O}{6} profile represent the absorption component kinematic subtypes we define in Section \ref{sec:types}.  `Broad' type \ion{O}{6} components are colored light green; `narrow' types are colored orange; and `no-low' types are colored purple. This color scheme will be consistent in all subsequent figures that include the component analysis. Note that all of the orange `narrow' type and green 'broad' type  \ion{O}{6} absorption features have low-ion matched counterparts, though not always in \ion{Si}{3} (\ion{Si}{3} is generally the best covered and most often detected low-ionization state transition). The purple `no-low' type  \ion{O}{6} absorption features do not have any low-ion counterparts. 
}
\label{fig:oxysillyprof}
\end{figure*}

\subsection{Profile Fitting and Qualitative Assessment}

 \cite{tumlinson11, werk13} and \cite{tumlinson13} summarized the COS-Halos Voigt profile fitting procedure for the  \ion{O}{6}, the low-ions, and the HI respectively.  We briefly repeat the relevant details. The procedure used to perform the fits and derive the column density $N$, Doppler $b$ parameter, and velocity offset $v$ for each component is an iterative fitting program that makes use of the  \verb1MPFIT1 software\footnote{http://cow.physics.wisc.edu/$\sim$craigm/idl/fitting.html}. The line profiles we derive from Voigt profile fitting are  convolved with the COS LSF as given at the nearest observed wavelength grid point in the compilation by \cite{coslsf}. Different transitions of the same ionic species are required to have the same component structure, and are therefore fit simultaneously to give a single solution. However, we do not impose such requirements on the different ionization states of the same element.   As we impose no restrictions on the component structure between different ionization states, any qualitative and quantitative similarities between the fits arise naturally.
 
  Following \cite[][see their Table 3]{tumlinson13}, we adopt $\pm$10 km s$^{-1}$ as the systematic uncertainty associated with the first-order and higher terms of the COS wavelength solution, and thus the {\emph{relative}} component velocity centroids and fitted $b$ values. When we consider the velocity centroids relative to the galaxy systemic redshift (typical systematic uncertainty $\sim$25 km s$^{-1}$, Werk et al. 2012), the root-square-sum gives a total error in the component velocity centroids of $\pm$30 km s$^{-1}$. Finally, we consider any line with a fitted $b$ value of $<$ 10 km s$^{-1}$ to be `unresolved'. 

Figure \ref{fig:highprof} shows COS-Halos continuum-normalized Voigt profile fits and corresponding data for 8 sightlines that cover \ion{O}{6} and at least one additional intermediate or high-ionization state line, either \ion{Si}{4}, \ion{C}{4}, or \ion{N}{5}.  A qualitative assessment of the overall agreement and component structure of the low-ion (e.g. \ion{Si}{3}), intermediate-ion (e.g.  \ion{Si}{4}) and high-ion ( \ovi) lines is complex. Generally,  there is good correspondence between the velocity ranges over which absorption is detected among the star-forming systems that exhibit \ovi\ absorption \citep{werk13}. 

Now turning to Figure \ref{fig:oxysillyprof}, we show the Voigt profile fits along the sightlines (within $\pm$ 300 km s$^{-1}$ of the galaxy systemic redshift) of all 24 star-forming COS-Halos galaxies that exhibit detections of  \ion{O}{6} (maroon). Where possible, we overlay \ion{Si}{3} profiles as solid black lines, and in all cases, we show the (often saturated) HI Ly$\alpha$ profile fits in light gray. In this profile comparison figure, we show the continuum normalized profiles without the spectral data to facilitate a simple by-eye comparison. There are impressive alignments in component structure (e.g. J1419$+$4207 132\_30),  and occasional misalignments between the strongest absorption components of each species (e.g. J1435$+$3604 68\_12), and sometimes \ion{O}{6} absorption without the presence of low-ion absorption (e.g. J1619$+$3342 113\_40).  This comparison again suggests the relationship between low and high ionization states of gas along the same COS-Halos sightlines is not straightforward. On a sightline by sightline basis, the relationship between the different ionization states can vary substantially. 


 Figure \ref{fig:oxysillyprof} displays a few cases in which individual components of \ion{O}{6}  do not appear to have {\emph{any}} affiliated low-ionization state metals. For example, J0914$+$2823 41\_27 (left-hand panel, third from bottom) contains one component at $-30$ km s$^{-1}$ that shows well-matched \ion{Si}{3} and broader, though aligned \ion{O}{6}, and another component at $-170$ km s$^{-1}$ that shows no \ion{Si}{3} nor any other low or intermediate-ionization state ion. This component at $-170$ km s$^{-1}$ {\emph{is coincident}} with a very broad component fit to Ly$\alpha$, with log N$_{\rm HI}$ = 14.07 cm$^{-2}$ and $b$ = 60 km s$^{-1}$. A similar situation exists for two high-velocity (v $>$ 150 km s$^{-1}$) components of  J1112$+$3539 236\_14. Finally, while J1435$+$3604 68\_12 may appear to follow this trend, we note that the narrower \ion{O}{6} component at $-60$ km s$^{-1}$ {\emph{does}} show a detection of \ion{C}{3} (saturated, and not shown), and is embedded in a DLA system. 

\subsection{Quantitative Comparison of \ion{O}{6} and \ion{Si}{3} Kinematics}

To directly compare the absorption profile characteristics of  \ion{O}{6} and other metal ions (e.g. \ion{Si}{3}), we must devise a quantitative method to match various absorption features along each line of sight.  We let the data drive our straightforward, observationally-motivated matching algorithm. As this analysis is focused on the highly-ionized gas, our starting point is the 39 individual fitted components of \ion{O}{6} along the 24 distinct sightlines within $\pm$300 km s$^{-1}$ of the galaxy systemic redshift. We attempt to match each of these individual \ion{O}{6} components with those of low and intermediate-ionization state species using their best-fitting velocity centroids from our Voigt profile analysis. Simply, we minimize the difference between the velocity centroids of \ion{O}{6} and other fitted metal ion components to identify matches.  We perform this test for every detected metal species and HI. Two distinct \ion{O}{6} components cannot have the same matching low or intermediate-ion component. The component with the best matching velocity centroid wins; the other component(s) is typically left unmatched (e.g. see the panel containing J1112$+$3539 236\_14 in Figure \ref{fig:oxysillyprof}; the purple dots mark the unmatched components). 

Additionally, we set a threshold for a metal-ion match such that the difference between the values of the velocity centroids cannot be greater than 35 km s$^{-1}$, a limit which likely captures the wavelength calibration uncertainty of the COS spectrograph over the full COS-Halos wavelength range. Further justifying this threshold is a natural break in the distribution of component velocity centroid differences of \ion{O}{6} and \ion{Si}{3} at 35 km s$^{-1}$. That is, there is no potential match that is rejected for a velocity centroid difference of 36 km s$^{-1}$; the next closest potential match beyond 35 km s$^{-1}$ lies at 65 km s$^{-1}$ (see the panel containing J1009$+$0713 204\_17 in Figure \ref{fig:oxysillyprof} for the unmatched \ion{O}{6} component at 75 km s$^{-1}$ marked by the purple dot, which is not matched with the \ion{Si}{3} component at 10 km s$^{-1}$). 

There are at least two major drawbacks to this method. First, matching components in this way leaves 12 low and intermediate-ion components behind because of the \ion{O}{6}-centric approach to matching. There are often more low-ion components than high-ion components along a given line of sight (e.g. see the panel containing J1016$+$4706 359\_16 in Figure \ref{fig:oxysillyprof}).  Second, there are likely some spurious associations owing to the underlying complex velocity fields of galaxy halos \citep[for details see][]{churchill15}, especially for those sightlines at low impact parameter that probe a huge range of galactocentric distances. By design, our observationally-motivated matching algorithm is well-matched to the data. This one-dimensional view,  with a velocity resolution of $\sim$15 km s$^{-1}$,  almost certainly does not fully capture the rich,  three-dimensional physics underlying the association between low and high-ion absorption.  However, the benefit of this  simple matching technique is that it allows us to quantify the general relationship between low and high ions discussed in the previous Section. It is our hope that being able to broadly quantify observed kinematic trends will serve to inform future physical models that aim to capture the multiphase ionization state of the CGM. 

  In total, there are 39 fitted \ion{O}{6} components in our 24-sightline sample. 31 of these components  can be matched with at least one low-ionization state component (e.g. \ion{Si}{3}) to within a velocity of 35 km s$^{-1}$.  There are 8 unmatched \ion{O}{6} components.   While we note that a careful inter-comparison of the \ion{H}{1} and the \ion{O}{6} could be fruitful \citep[e.g.,][]{savage11}, it is complicated by the strong, often-saturated, and complex associated \ion{H}{1} absorption \citep{tumlinson13} that may be tracing a cooler, more neutral gas phase. 
    
\begin{figure}[h]
\begin{centering}
\hspace{0.1in}

\includegraphics[width=0.85\linewidth]{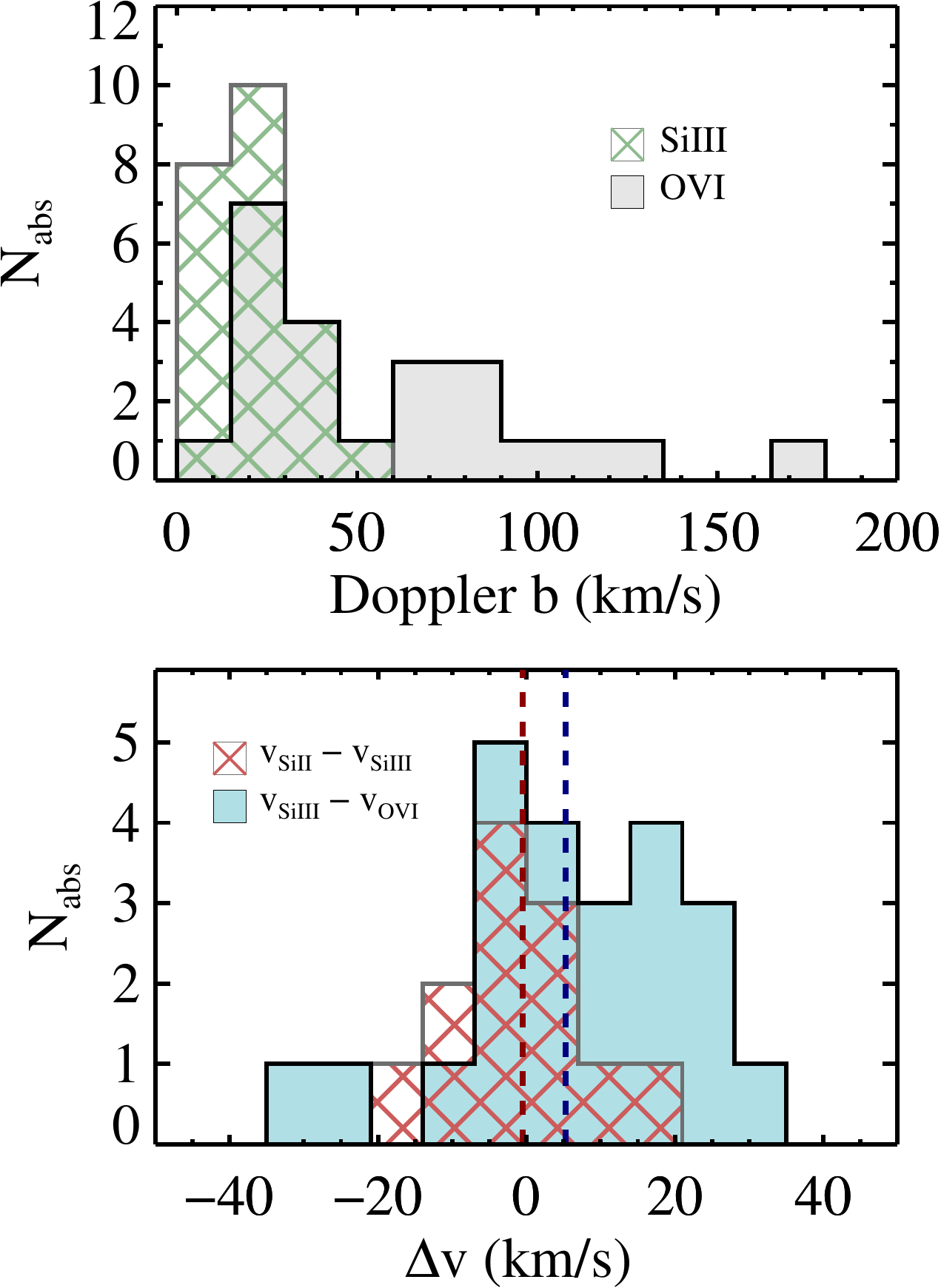}
\end{centering}
\caption{ Upper panel: the distribution of Doppler $b$ parameter values for matched components of \ion{Si}{3} and \ion{O}{6}. \ion{Si}{3} $b$ values are shown in the green crosshatched histogram (mean $b$ = 22 km s$^{-1}$) and \ion{O}{6} $b$ values lie within the gray-filled histogram (mean $b$ = 57 km s$^{-1}$). Lower panel: the velocity centroid differences between matched \ion{Si}{3} and \ion{O}{6} absorption components (solid blue) and matched \ion{Si}{2} and \ion{Si}{3} components (crosshatched orange). The mean difference between the central velocities of  \ion{Si}{3} and \ion{O}{6} absorption is 5 km s$^{-1}$, marked by the dark blue dashed line, indicating \ion{O}{6} is typically found at lower velocities than \ion{Si}{3}, though this is a minor effect. Comparing this $\Delta$v to that of  \ion{Si}{2} and \ion{Si}{3}, we find the latter to be generally better matched in velocity space, with a mean of 0 km s$^{-1}$. For both panels, we match components based on a simple algorithm described in Section \ref{sec:types} that minimizes differences between central velocities of fitted components. 
}
\label{fig:bvhist}
\end{figure}

  Figure \ref{fig:bvhist} shows two histograms that compare the matched absorption components of \ion{Si}{3} and \ion{O}{6}, the two most commonly detected metal species in the COS-Halos dataset. The two relevant absorption profile characteristics to compare are the Doppler $b$ parameters for each ion (upper panel), and the velocity centroid differences between the matched components (lower panel). The typical errors in the $b$ parameters from our Voigt profile analysis are $\approx$ $\pm$20\%. The typical errors in the fitted velocity centroids are $\pm$ 7 km s$^{-1}$, and thus the errors in the velocity centroids are dominated by the systematic uncertainty in the COS wavelength calibration $\pm$10 km s$^{-1}$. We note that these histograms exclude a subset of absorption-line components that cannot be matched following our procedure outlined above (8/39). We will consider this subset of the general population of absorbers separately. 

In the upper panel of Figure \ref{fig:bvhist}, we show the distribution of Doppler $b$ parameters for the 23 matched \ion{Si}{3} and \ion{O}{6} components separately. The difference between the two ions is immediately apparent by examining the distributions of their Doppler line widths.  A two-sample KS test,  using the maximum deviation of 0.58, yields a probability of 0.00017 that the two $b$-value distributions are drawn from the same distribution.  The mean line width for \ion{Si}{3} is 22 km s$^{-1}$, a value typical of the low-ionization state gas in COS-Halos and other absorption surveys, and comparable to the velocity resolution of COS \citep[e.g.][]{savage14, lehner14}. In contrast, the mean line width for \ion{O}{6} is 55 km s$^{-1}$, a value that is considerably broader than the \ion{O}{6} absorption in general absorption-line centric surveys \citep[discussed in Section \ref{sec:ovicomparisons}; ][]{tripp08, muzahid12, savage14, danforth16}

Turning to the lower panel of Figure \ref{fig:bvhist}, we directly examine the differences between the velocity centroids of the matched components of \ion{Si}{3} and \ion{O}{6}. For reference, we also show the velocity centroid differences between \ion{Si}{2} and \ion{Si}{3}. The average offset (absolute value; not shown) between our matched SiIII and \ion{O}{6} is only 18 km s$^{-1}$.  For the low-ions, the average offset between \ion{Si}{2} and \ion{Si}{3} (or \ion{C}{2} and \ion{C}{3}) is 7 km s$^{-1}$, with much greater overall agreement, and consistent with the errors in the relative wavelength scale of COS. We additionally mark the mean values of the velocity centroid differences for both pairs of ions on the lower panel of Figure \ref{fig:bvhist}. For offsets between velocity centroids that arise due to uncertainties in the wavelength scale and velocity resolution limitations, we expect this mean velocity offset between components to be 0 km s$^{-1}$, as it is for \ion{Si}{2} and \ion{Si}{3}. The 5$\pm$3 km s$^{-1}$ mean value for the offsets between \ion{Si}{3} and \ion{O}{6} is only marginally different from no offset at all, yet may hint that \ion{O}{6} sits at slightly smaller $\Delta$v (relative to galaxy systemic). Such an offset may be expected under certain physical conditions (e.g. cooling flow models) that we will explore in later sections. 

For reference, two sets of unassociated absorbers distributed randomly within $\pm$ 250 km s$^{-1}$ would show the type of agreement between their velocity centroids seen in Figure \ref{fig:bvhist} only 4 $-$ 8\% of the time.  A two-sample KS-test between a random distribution of velocity centroid differences within $\pm$ 250 km s$^{-1}$  and the observed distribution derived from our matching algorithm shown in the lower panel of Figure \ref{fig:bvhist} yields a probability of 0.0000013 that our observed component overlap is random.  The association between low and high ion absorber velocity centroids is significant at a level of $\approx$ 5$\sigma$. Furthermore, the mean difference in $b$ values for matched components of \ion{Si}{3} and \ion{O}{6} is 40 km s$^{-1}$. We further note that $\sim$45\% of matched components of the two ions show differences in $b$ values $<$ 15 km s$^{-1}$ and Doppler $b$ values $<$ 25 km s$^{-1}$. The $b$ value distributions shown in the upper panel of Figure \ref{fig:bvhist} reflect this dichotomy between CGM absorbers that show coincident low and high-ion absorption. 

\subsubsection{Identifying Three Kinematic Subtypes}
\label{sec:types}

Our analysis of matched and unmatched components of \ion{O}{6} reveals three distinct kinematic types of absorption present in the halos of star-forming L$^*$ galaxies. Any physical picture of the CGM must incorporate all three distinct kinematic types of absorption, and is therefore likely to require multiple physical origins for \ion{O}{6}. 
\begin{description}
\item{{\bf{Broad:}} Low-ion matched \ion{O}{6} absorption with $b$ $>$ 40 km s$^{-1}$ appears to represent 16/39 (41\%) of the components. These absorbers have velocity centroids that correspond well with low and intermediate ionization state gas, but exhibit significantly broader line widths,  $\Delta b$ $>$ 30 km s$^{-1}$. Accordingly, we will refer to this kinematic type as `broad' in subsequent text and figures; on figures it will be represented by light green squares. }
\item{{\bf{Narrow:}} Low-ion matched \ion{O}{6} absorption with $b$ $<$ 35 km s$^{-1}$  represents 15/39 (38\%) of \ion{O}{6} absorbers, and is characterized by well-aligned low-ionization state components with similar line widths,  $\Delta$b $<$ 15 km s$^{-1}$. We will refer to this kinematic type as `narrow' in subsequent text and figures; on figures it will be represented by orange diamonds. }
\item{{\bf{No-low:}} Finally, this third type of absorption occurs in 21\% of \ion{O}{6} absorbers (8/39), and indicates a non-detection of low or intermediate ionization state gas within $\sim$50 km s$^{-1}$. These absorption components are typically broad, with mean Doppler $b$ parameters of $\sim$ 50 km s$^{-1}$ (though two have $b$ $<$ 30 km s$^{-1}$), and tend to have associated HI (within 50 km s$^{-1}$) with log N$_{\rm HI}$ $\approx$ 13.4 - 15 cm$^{-2}$ and mean $b_{\rm HI}$  = 40 km s$^{-1}$. We refer to this kinematic type as `no-low' in subsequent text and figures; on figures this type will appear as purple circles. }
\end{description}

We emphasize that these definitions are based purely on absorption-line kinematics and do not explicitly select for any galaxy property or absorption line column density. However, there are two possibly significant selection effects that should serve as caveats to this statement. The `broad' type absorption components matched with low-ion absorption components tend to select for the highest \ion{O}{6} column density components. At the velocity resolution of COS, and the S/N of the COS-Halos data, the blending of narrow components separated by small offsets  in velocity space (e.g. $<$ 10 km s$^{-1}$) is entirely possible. Thus, by selecting for broad absorption features, we could be preferentially picking components with many narrow components that contribute to the total column density. We explore this possible selection effect further in the next section.  Second, that the `no-low' absorbers are the least common in galaxy halos is not surprising. This type tends to consist of lower \ion{O}{6} column density absorbers with broad line widths.  Thus,  its signal can easily be drowned out by the relatively stronger signals from `broad' and `narrow' matched components along the same lines of sight. The effect is similar to that which is pointed out by \cite{zheng15}, in which they report the Milky Way CGM is `half-hidden' by strong absorption from the ISM at  $-100$ km s$^{-1}$ $<$ v$_{\rm LSR}$ $<$ 100 km s$^{-1}$. We explore this effect further in Section \ref{sec:galprop}. 

\subsubsection{Comparison between CGM and IGM \ion{O}{6} Absorption}
\label{sec:ovicomparisons}

There is extensive literature on the incidence of \ion{O}{6} gas along
extragalactic sightlines and its association with galaxies and
large-scale structures \citep[e.g.][]{tripp08,wakker09,prochaska11,stocke13,savage14}.
These `intervening'  \ion{O}{6} absorbers found along blindly-selected QSO sightlines have lower
column densities (log N$_{\rm OVI}$ $\approx$ 13.8; Danforth et al. 2016) than the \ion{O}{6} gas seen in COS-Halos (log N$_{\rm OVI}$ $\approx$ 14.8).  Statistically, the majority of the blindly-selected intervening \ion{O}{6} absorbers must occur far from \lstar\ galaxies \citep{prochaska11}, either around lower mass galaxies or in the IGM itself.  One may consider, therefore, whether this gas constitutes a qualitatively different population with a unique physical origin.

\begin{figure}[t!]
\begin{centering}

\includegraphics[width=0.95\linewidth]{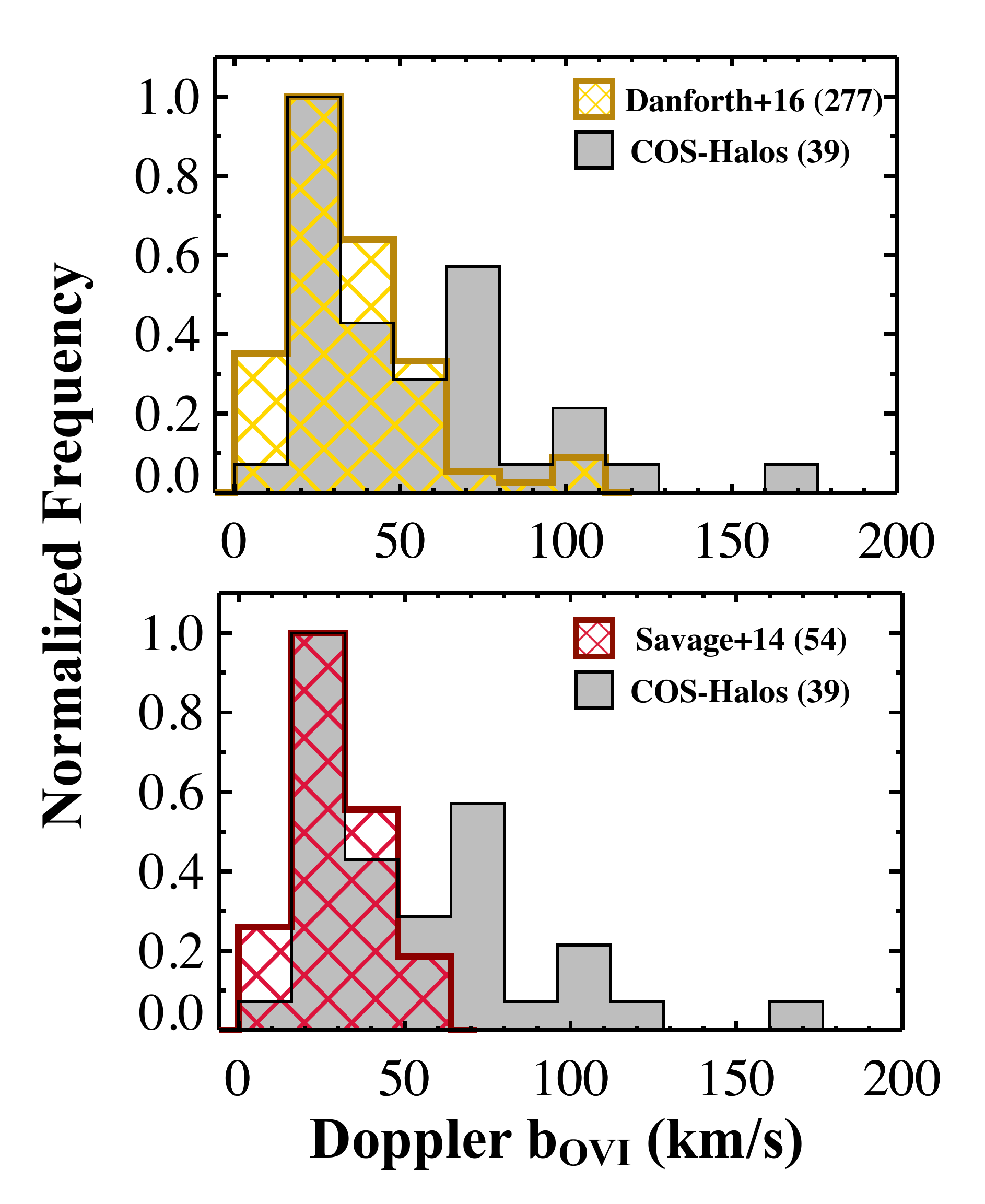}
\end{centering}
\caption{ Distribution of the Doppler $b$ parameter values for the 39  \ion{O}{6} components in the CGM of COS-Halos SF galaxies (gray histogram) compared to those of the low-z IGM absorbers studied in Danforth et al. (2016; D16, crosshatched yellow histogram, top panel) and Savage et al. (2014; S14, crosshatched red histogram, bottom panel). The histograms have been normalized such that the peak of each distribution is 1. The D16 distribution includes $b$ parameters from 277  \ion{O}{6} components with a mean $b$ parameter of 35 km s$^{-1}$ while the S14 distribution includes 54 \ion{O}{6} absorber components that are well-aligned with \ion{H}{1}, having a mean $b$-value of 30 km s$^{-1}$. In contrast, the mean of the COS-Halos \ion{O}{6} linewidths is 55 km s$^{-1}$. 
}
\label{fig:bhists}
\end{figure}

Toward this end, we show a comparison of the distributions of Doppler $b$ parameters from two studies of the low-redshift (z $<$ 0.9) IGM \citep{savage14, danforth16} with that of COS-Halos in Figure \ref{fig:bhists}. Throughout this section,  we will refer to \cite{danforth16} as D16 (crosshatched yellow histogram) and \cite{savage14} as S14 (crosshatched red histogram).  All three studies detect \ion{O}{6} in HST/COS G130M or G160M  spectra, and therefore have similar spectral resolution.  The COS-Halos data (gray histogram) show a significantly larger fraction of broad \ion{O}{6} components, with a mean value of the Doppler $b$ parameter of 55 km s$^{-1}$. Approximately 60\% of the COS-Halos \ion{O}{6} line profiles have $b$ $>$ 35 km s$^{-1}$, are symmetric, and have little internal structure, despite often aligning well with the more highly-structured low-ionization state absorption.  Furthermore, both S14 and D16 find more narrow \ion{O}{6} lines than COS-Halos. A two-sided KS test between COS-Halos, S14, and D16, respectively, yields probabilities of the null hypothesis of 0.00041 and 0.0018.  These differences between the samples could occur if COS-Halos \ion{O}{6} profiles consist of many blended components, such that narrow components are increasingly difficult to recognize.  Or, they may indicate a physically distinct origin for CGM and IGM \ion{O}{6}.  

The study by D16 includes HST/COS data of 82 UV-bright QSOs containing 280 \ion{O}{6} systems with 0.1 $<$ z $<$ 0.74 (277 \ion{O}{6} $\lambda$1031 \AA~ components shown; crosshatched yellow histogram). The mean of \ion{O}{6} $b$-parameters included in D16 is 35 km s$^{-1}$, with a low fraction of absorbers exhibiting values $>$ 60 km s$^{-1}$.  Similarly, S14 find a mean value of $b$ for \ion{O}{6} to be 29 km s$^{-1}$ in 14 high S/N QSO spectra taken with COS (crosshatched red histogram).   Although not shown on Figure \ref{fig:bhists}, we also note that  \cite{muzahid12} find a mean value of $b$ for high-z (1.9 $<$ z $<$ 3.1) IGM \ion{O}{6} absorbers to be 28 km s$^{-1}$, and using higher resolution VLT UVES data. Using STIS E140M data,  \cite{tripp08} find a mean \ion{O}{6} $b$-value of 27 km s$^{-1}$ for 77 intervening absorbers at z $<$ 0.5, with only 2 out of 77 \ion{O}{6} components with $b$ $>$ 60 km s$^{-1}$ (see their Figure 13).


We caution that such comparisons of fitted absorption-line widths in spectra having different S/N and spectral resolution are subject to significant systematics.  Our direct comparison between studies using COS spectra in Figure \ref{fig:bhists} eliminates the second source of bias, but the mismatch between the S/N of the various studies could impact the $b$ parameter distributions. The S/N of the COS-Halos data considered here is considerably lower ($\sim$10) than both D16 (S/N = 15 - 78) and S14 (S/N = 16 - 155). It is therefore possible higher quality spectra would reveal that weaker, narrower components comprise the very broad \ion{O}{6} we find in our data. Nonetheless, we find it unlikely that the difference in S/N is driving the difference between b-value distributions of the inner CGM and general IGM. In particular, a direct comparison between the \ion{Si}{3}  $b$-values included in the D16 study and those of COS-Halos give very good agreement: 25 km s$^{-1}$ vs. 22 km s$^{-1}$ respectively, even though  \ion{Si}{3} would be subject to the same systematics as \ion{O}{6}. Thus, despite the above caveats, we conclude that {\emph{ very broad \ion{O}{6} absorption is an important characteristic of  the CGM of star-forming L$^*$ galaxies.}} 

\subsubsection{Physical Insights from \ion{O}{6} Absorption Line Widths}
\label{sec:heckman}

\begin{figure*}[t!]
\vspace{0.1in}
\begin{centering}
\hspace{1.2in}
\includegraphics[width=0.60\linewidth]{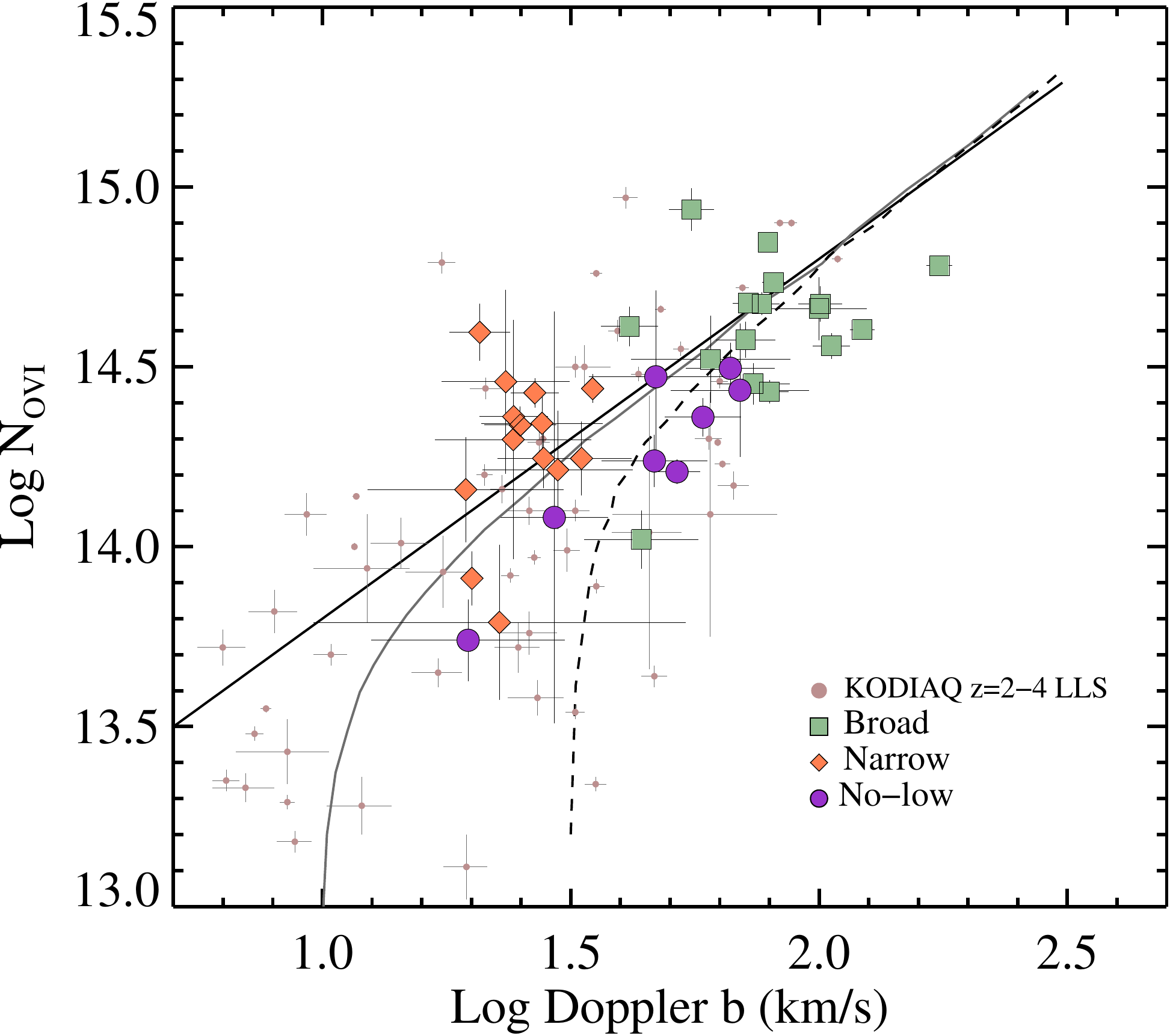}
\end{centering}
\caption{ \ion{O}{6} column density versus Doppler $b$-parameters for the individual components colored by type: green is `broad' type; orange is `narrow' type; and purple is `no-low' type. For comparison, we show the `robust' estimates of the same parameters taken from the KODIAQ sample of z$=$2$-$4 Lyman limit systems as small light pink data points \citep{lehner14}. Their data were obtained with Keck/HIRES. The two curves show the expectations of N$-$b behavior for radiatively cooling, collisionally-ionized gas for  two temperatures, T$_{\rm OVI}$ =  10$^{5}$ K (black-dashed) and  T$_{\rm OVI}$ =  10$^{6}$ K (solid gray) \citep{heckman02}.  The curves are calculated assuming that the cooling flow velocity is equivalent to the b-value. The solid black straight line shows the effect of blending multiple narrow components with 0 km s$^{-1}$ velocity offsets.   }
\label{fig:Nb}
\end{figure*}

In principle, the absorption line widths, quantified by the Doppler $b$ parameter, encode physical information about the absorption. For example, one can place an upper limit on the gas temperature, assuming thermal line broadening, using the relation: \begin{equation} b_{\rm th} = \left(\frac{2kT}{m}\right)^{\frac{1}{2}}. \end{equation} Here, $b_{\rm th}$ is the thermal Doppler $b$ parameter, $k$ is the Boltzmann constant, $T$ is the gas temperature, and $m$ is the atomic mass. The typical velocity dispersions of the absorption components are small for the low-ionization state gas, with typical Doppler $b$ values ranging from 5 km s$^{-1}$ to 20 km s$^{-1}$. This range of $b$ values implies the low-ion lines are marginally resolved or unresolved, and trace gas temperatures  $\lesssim$10$^{5}$ K, consistent with a plasma photoionized by an incident radiation field. The Doppler $b$ values of the \ovi~absorption lines are systematically higher for absorption components centered at the same redshift. This discrepancy in absorption line widths may indicate that \ovi~traces gas with a higher temperature, limited to $T \lesssim10^{6.5}$ K. On the other hand, it may require that turbulence is dynamically significant for  \ion{O}{6} in galaxy halos \citep[see][for a detailed discussion]{faerman16}, or that \ion{O}{6} spans a significantly larger fraction of a complex halo velocity field than the low-ion absorption lines \citep{stern16}. 

We can roughly quantify the non-thermal component of the line width under the assumption of a characteristic temperature for \ion{O}{6}-bearing gas, where:  \begin{equation} b^{2} = b^{2}_{\rm th} + b^{2}_{\rm nt}. \end{equation} We proceed under the caveat noted previously regarding the possible blending of narrow components with velocity offsets $<$ 10 km s$^{-1}$ that can contribute significantly to the measured line widths. The measured line widths are also impacted by instrumental broadening, though we have tried to account for this as well as possible in our Voigt profile fitting procedure. As such, the non-thermal values reported here should be considered upper limits. Photoionization equilibrium models for optically thin gas give an equilibrium gas temperature, T$_{\rm PIE}$ $\approx$ 10$^{4.5}$ K \citep[e.g.][]{ferland13}. Collisional ionization equilibrium models give T$_{\rm CIE}$ $\approx$10$^{5.4}$ K where the abundance of \ion{O}{6} peaks \citep[e.g.][]{gs07, oppenheimer13}. Thus, the expected line widths of \ion{O}{6} in each case due to only thermal broadening are 6.4 km s$^{-1}$ and 16.2 km s$^{-1}$, respectively. Both values are significantly lower than the majority of our measured $b$ values for \ion{O}{6}.  The average non-thermal contribution to the \ion{O}{6} line widths is $\sim$40 - 50 km s$^{-1}$.  We note for \ion{Si}{3} at T$_{\rm PIE}$ = 10$^{4.3}$ K, the expected line width due to thermal broadening alone is 3.4 km s$^{-1}$. Thus, the non-thermal contribution to the \ion{Si}{3} line widths is limited to be $\lesssim$ 20 km s$^{-1}$. 

\cite{heckman02} analyzed the relationship between \ion{O}{6} column density and absorption-line width for a wide range of physically diverse environments, incorporating measurements from the Milky Way HVCs, Magellanic Clouds, starburst galaxies, and the IGM. They found a relationship consistent with that predicted theoretically  for a radiatively cooling flow of hot gas as it passes through the coronal temperature regime \citep{edgar86}, shown as the dashed and solid gray lines here in our Figure \ref{fig:Nb}. \cite{wakker12} describe in detail the radiative cooling flow model first proposed by \cite{shapiroandmoore76} and later developed by \cite{edgar86, shapiro91, benjamin94}.  This model allows gas to cool via collisions as it moves at a constant speed, and predicts  the relation shown in Figure \ref{fig:Nb}. Heckman and others \citep[e.g.][]{bordoloi16} have argued that this relationship, which spans orders of magnitude in column density,  indicates a generic origin of \ion{O}{6} in collisionally-ionized gas.  As we discuss in Section \ref{sec:analysis}, the authors note that radiatively cooling gas behind fast shocks is not the only physical model that can lead to the relationship between $\novi$ and $b$ \citep[see also:][]{od09}. For example, turbulent mixing can give rise to such a relation since the inflow rate of mass is explicitly tied to the cooling rate in these models \citep[e.g.][]{kwak11}. 

 In the 14 years following the result first shown by \cite{heckman02}, there have been many others to investigate this correlation (or in some cases, its absence) extending it to stronger starbursting galaxies \citep{grimes09}, many more z $<$ 0.5 intervening IGM absorbers \citep{tripp08}, z$= 2-4$ Lyman limit systems \citep{lehner14}, and incorporating modern measurements for the CGM and IGM \citep[including COS-Halos;][]{bordoloi16}. We now explore this trend seen in the COS-Halos \ion{O}{6} absorbers in Figure \ref{fig:Nb}, as a function of their kinematic subtype. We show the high-z LLS absorbers included in the KODIAQ sample \citep{lehner14} for reference. 

The first thing to notice about the data plotted in Figure \ref{fig:Nb} is that `narrow' and `broad' type \ion{O}{6} absorbers separate nicely on the $b$ parameter axis. Of course, this is by design (see Section \ref{sec:types}). Furthermore,  each kinematic subtype traces a distinct region of the parameter space, with one or two exceptions. This facet of the figure is not explicitly by design, though selection effects may be playing some role in this separation. 
Second, it appears as though the \ion{O}{6} absorbers indeed follow the trends delineated by the cooling flow curves. A Spearman-Rho test on the full sample indicates a 3.5 sigma correlation in the N$_{\rm OVI}$ vs. $b$ parameter space, with a rank coefficient of 0.70. The `no-low' type absorbers trace the lower edge of the N-b correlation seen in the data. 

Now, we consider the above results in the context of the Heckman model. The solid black line shows the the effect of blending multiple components with negligible velocity offsets, which, at the \ion{O}{6} column densities probed by COS-Halos, is nearly indistinguishable from the gray line, which marks gas radiatively cooling at T$_{\rm OVI}$ = 10$^{6}$ K. For increasing velocity offsets of the narrow, unresolved components, this solid black line moves down and to the right (see Heckman et al. 2002) and becomes even more consistent with the Heckman model predictions.  For this reason, we cannot consider a priori the observed correlation to be a smoking gun for collisional ionization. 

If component blending is indeed dominating the trend seen in Figure \ref{fig:Nb}, we may expect to see such a correlation for the low-ionization state absorption components that are matched to the \ion{O}{6} components. There is no such correlation for {\emph{any}} of the low or intermediate-ions, neither for only the matched components nor all low-ion components regardless of their \ion{O}{6} correspondence. Thus, we find it unlikely that the N-b correlation in Figure \ref{fig:Nb} is due to blending alone. 

Finally, we point out that the `narrow' type absorbers (orange diamonds) are actually inconsistent with the Heckman models to a significant degree. At the line-widths tracked by the `narrow' lines, cooling flow models tracing 10$^{5-6}$ K gas predict N$_{\rm OVI}$ values $<$ 10$^{14}$ cm$^{-2}$, while 87\% (13/15) of `narrow' absorbers show N$_{\rm OVI}$ in excess of this value. Considered alone, `narrow' type \ion{O}{6} absorbers do not exhibit any correlation between their column densities and line-widths. The `no-low' absorbers (purple circles) are most consistent with the Heckman models,  lie below the blending line, and track a rather tight, 2.5$\sigma$ correlation between their column densities and line widths (the Spearman-Rho coeffcient $>$ 0.70), despite having only 8 data points. There are several other additional factors (some explored below) including broad associated \ion{H}{1} and a complete lack of low-ionization state gas that make it likely this `no-low' kinematic subtype represents gas cooling via collisional ionization. The `broad' absorbers (green squares) alone only track a 1.6$\sigma$ correlation as indicated by a Spearman-Rho test, and exhibit a  considerable degree of scatter. Furthermore, as previously mentioned, this type of absorber is most prone to blending effects.

\subsubsection{Correlating Kinematics with Galaxy Properties}

\label{sec:galprop}

\begin{figure}[h]
\vspace{0.1in}
\begin{centering}
\hspace{-0.2in}
\includegraphics[width=1.0\linewidth]{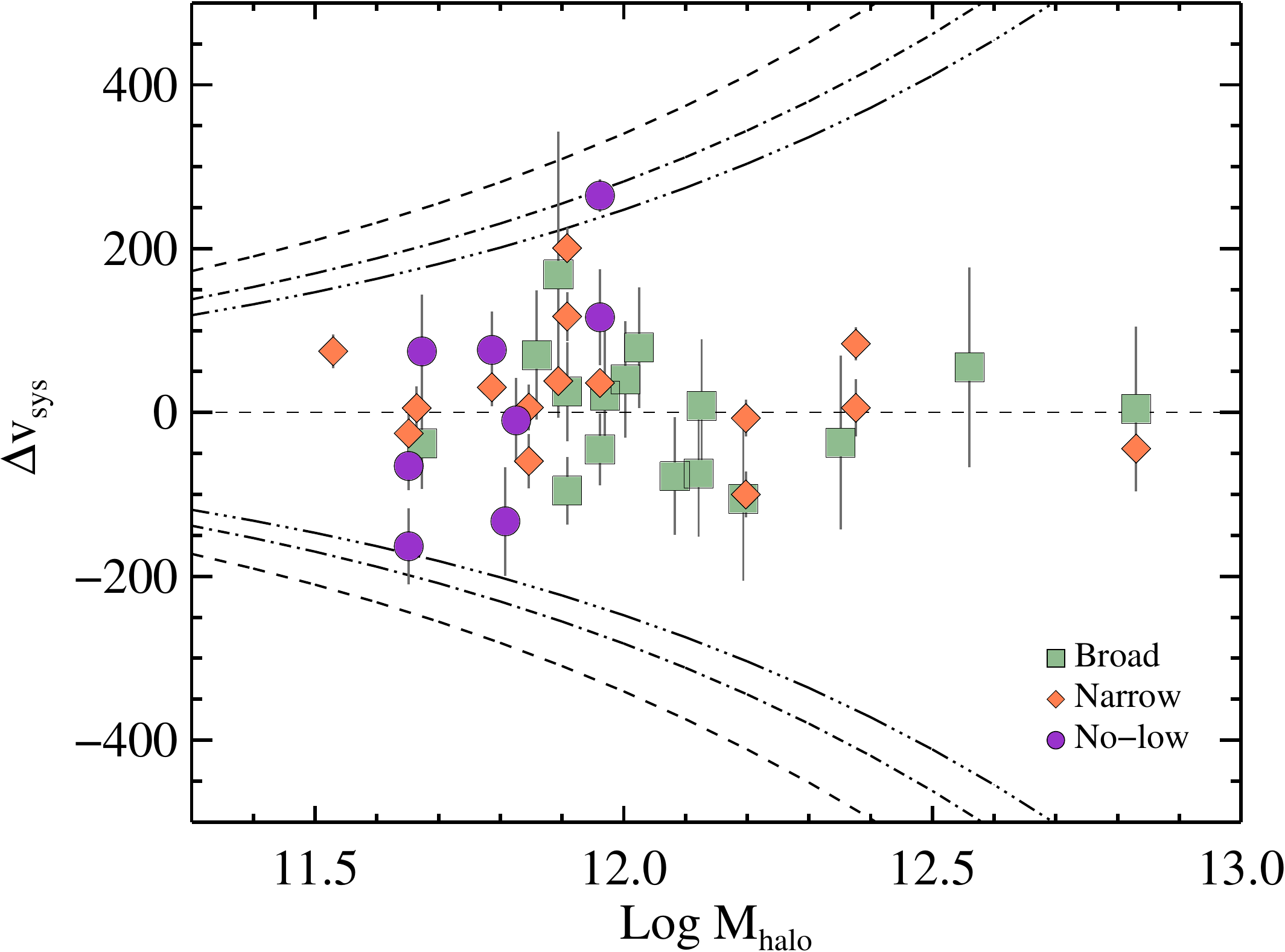}
\end{centering}
\caption{ The difference between the galaxy systemic redshift (the assumed `rest' frame) and the velocity centroids of the 39 \ion{O}{6} components in the 24 star-forming COS-Halos galaxy sub-sample addressed in this work, as a function of the galaxy halo mass, determined by abundance matching \citep{werk12, tumlinson13}. As in previous figures, the data points are colored by kinematic subtype: green squares indicate `broad' type \ion{O}{6} absorption; orange diamonds show `narrow' type absorption; and purple circles mark the `no-low' type absorption. The vertical gray lines track the line-widths of the components, with a total length given by $\pm$ $b_{\rm Doppler}$. From the inside moving outward, the curves trace the halo escape velocities from 50, 100, and 150 kpc, respectively. }
\label{fig:vesc}
\end{figure}

The primary benefit of relying upon the COS-Halos dataset is the well-characterized host galaxy sample \citep{werk12}, and the resulting opportunity to relate the properties of the gas 10 $-$150 kpc from a galaxy to the galaxy stellar and halo masses and global star-formation rates (SFRs). Toward this end, Figure \ref{fig:vesc} shows the difference between the galaxy systemic redshift (the assumed `rest' frame) and the velocity centroids of the \ion{O}{6} components ($\Delta$v$_{\rm sys}$) on the y-axis, as a function of the galaxy halo mass, determined by abundance matching \citep{werk12, tumlinson13}. To determine the `halo' masses, as previously described by \citep{werk14}, at a galaxy stellar mass given by $kcorrect$ \citep{kcorrect} from the SDSS $ugriz$ photometry, we interpolate along the halo abundance matching relation of \cite{moster10}.  

It is immediately apparent from Figure \ref{fig:vesc} that the broad \ion{O}{6} absorbers with no corresponding low-ionization state gas (`no-low'; purple circles) are absent in galaxies with log M$_{\rm halo}$ $>$ 12.0. A two-sample KS test on the `no-low' M$_{\rm halo}$ distribution vs that of the `broad' and `narrow' types indicates this absence is statistically significant at the 3$\sigma$ level with a Kolmogorov-Smirnov statistic of 0.6. Furthermore, these kinematic subtypes were selected without regard to galaxy properties, and there is no plausible selection effect that could account for the absence of \ion{O}{6} absorbers in the star-forming galaxies with more massive halos. We do note that we are likely to miss a significant fraction of `no-low' absorbers potentially washed out by strong `broad' and `narrow' absorbers along the same sightline, however that will bias the impact parameter distribution of the kinematic subtypes, as we discuss later in this section.  COS-Halos evenly samples the full range of halo masses in impact parameter space for the 24 galaxies considered. 

The distribution of $\Delta$v$_{\rm sys}$ compared to the halo escape velocities is the other striking feature of Figure \ref{fig:vesc}. For both `broad' and `narrow' type absorbers coincident with low-ionization state gas, the $\Delta$v values are rather tightly concentrated around 0 km s$^{-1}$ as compared to the allowed range of $\Delta$v for bound gas especially for log M$_{\rm halo}$ $>$ 12. In contrast, the `no-low' type absorbers occupy the full range of allowed $\Delta$v for their given range of M$_{\rm halo}$.  As we will discuss in Section \ref{sec:discussion}, the virial temperature for 11.5 $<$ log M$_{\rm halo}$ $<$ 12.0 is 10$^{5.3 - 5.5}$ K, {\emph{exactly}} the temperature at which the \ion{O}{6} gas fraction peaks in collisional ionization models, both in and out of equilibrium \citep[e.g.][see their Equation 1]{oppenheimer16}. We will argue that the abrupt truncation of the halo mass distribution at log M$_{\rm halo}$ $<$ 12.0, and  the statistically significant correlation between log N$_{\rm OVI}$ and $b$ values for the `no-low' type absorbers are strong pieces of evidence that this kinematic subtype is tracing halo gas at T$\approx$T$_{\rm vir}$, cooling via collisions. 

\begin{figure}[t!]
\begin{centering}
\includegraphics[width=0.45\textwidth]{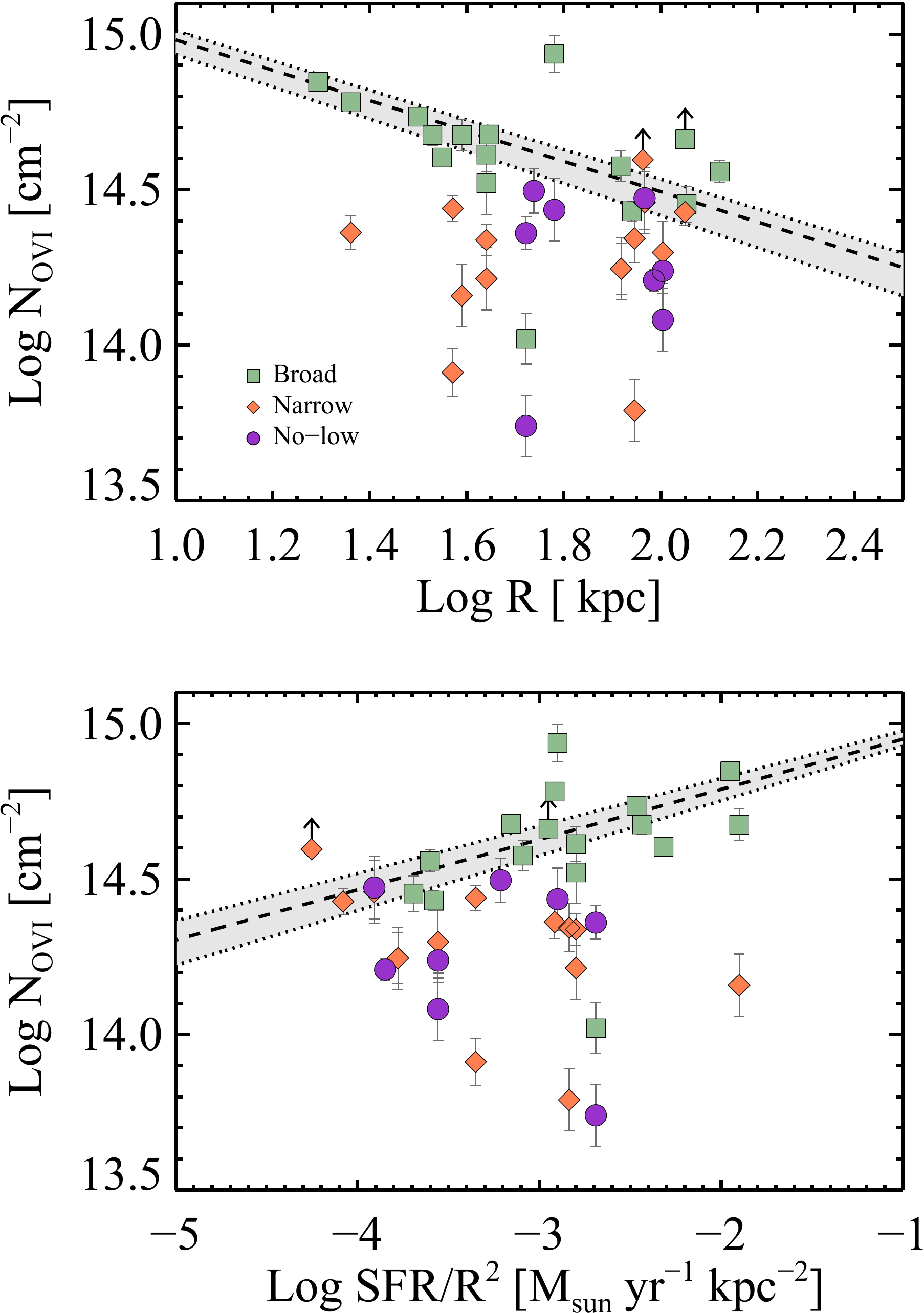}
\end{centering}
\caption{ Top: log  $\novi$ as a function of impact parameter, R, for the individual \ion{O}{6} components colored by type where green squares indicate `broad' type \ion{O}{6} absorption; orange diamonds show `narrow' type absorption; and purple circles mark the `no-low' type absorption. The 2.9 $\sigma$ correlation between the total line of sight AODM column densities and impact parameter is shown as a dashed line with 1$\sigma$ uncertainty marked by the gray shaded area.  The `broad' type absorbers dominate and define this trend, while the other two types exhibit no correlation between log  $\novi$ and R. Furthermore, the `no-low' type absorbers (purple circles) lie at impact parameters R $>$ 50 kpc, while the other two types span the full range of impact parameters. Bottom: Similar to the top plot, but now the x-axis shows  the galaxy SFR, scaled inversely by the impact parameter squared.  The 2.7$\sigma$ correlation for the total $\novi$ for each galaxy halo (i.e. not by component) is shown in light gray. The `broad' type absorption in each galaxy halo is driving the overall correlation between log  $\novi$ and SFR/R$^{2}$.  The other two types are not correlated at all with SFR/R$^{2}$ (or SFR).  }
\label{fig:sfrovi}
\end{figure}

We turn now to the top panel of Figure \ref{fig:sfrovi}, which shows N$_{\rm OVI}$ as a function of impact parameter, R, with data colored by kinematic subtype. The global COS-Halos trend for star-forming galaxies, which uses the total integrated N$_{\rm OVI}$ within $\pm$300 km s$^{-1}$ of galaxy systemic,  is shown by the dashed black line with 1$\sigma$ uncertainty marked by the gray shaded region for reference. As discussed previously, the `no-low' type absorbers are found only at R $>$ 50 kpc, possibly due to selection effects that cause the low-column density broad absorption lines to be blended with `narrow' and `broad' type absorbers along the longer total path lengths probed at lower impact parameters. Additionally, we point out that the `broad' type \ion{O}{6} absorbers with matched low-ion counterparts are setting the global COS-Halos trend between N$_{\rm OVI}$ and R. This is unsurprising given the previous observation that these absorbers dominate the total column density of the SF galaxies. Yet it is significant; Figure \ref{fig:sfrovi} leads us to conclude that `broad' type absorbers drive the global correlations between \ion{O}{6} and galaxy properties.

Perhaps the most perplexing trend apparent in the full COS-Halos sample is the correlation between the global galaxy SFR and the total integrated \ion{O}{6} column density \citep{tumlinson11}.  Particularly significant is the strong presence of \ion{O}{6} absorption around star-forming galaxies and its rarity around galaxies with no detectable level of star-formation. This dichotomy may imply that the material bearing \ion{O}{6} is somehow transformed or lost from the CGM once the galaxy's star formation quenches \citep[but see:][]{oppenheimer16}.  For the star-forming galaxies in consideration here, galaxy SFR and $\novi$ are not significantly correlated (40\% chance a correlation is not present), though we do note that the significance of this correlation increases to 3$\sigma$ (0.1\% chance a correlation is not present) when we consider the upper limits to the SFR and $\novi$ implied by the spectroscopic observations of the 16 passive galaxies in the COS-Halos sample. 

In Section \ref{sec:analysis}, we will explore the evidence for (and rather dramatic implications of)  \ion{O}{6} being produced in part by ionizing radiation either indirectly or directly due to the galaxy's ongoing SFR. If this were true,  we  might expect its strength to correlate directly with SFR/R$^{2}$, which is explored in the bottom panel of Figure \ref{fig:sfrovi}. Additionally, we note that other phenomena, such as starburst driven winds, may also result in such a correlation. On this figure, we show the 2.7$\sigma$ correlation\footnote{More specifically, there is a 0.8\% chance a correlation is not present; according to a Kendall-Tau test including measurement errors and censoring by lower limits.} for the total integrated N$_{\rm OVI}$ as a function of SFR/R$^{2}$ as the dashed black line with the 1$\sigma$ errors enclosed in the gray shaded area. The  power law fit with 1$\sigma$ errors is: 
\begin{equation} \novi~[\rm cm ^{-2}] = 10^{15.1\pm0.17} \left(\frac{\rm SFR}{\rm R^2}\right)^{0.16\pm0.05}\end{equation} 
where SFR is in M$_{\odot}$ yr $^{-1}$ and impact parameter, R, is in kpc. We note that this correlation is driven primarily by the decrease in $\novi$ with impact parameter, and could just as easily represent a gas surface density gradient in the galaxy halo. Furthermore, neither the `narrow' nor `no-low' absorber column densities are correlated with SFR/R$^{2}$. 

\section{Analysis of Ionization Processes}
\label{sec:analysis}

We now explore the various ionization processes that may produce a highly ionized
plasma bearing \ovi.  We then test whether these
scenarios are consistent with the observed column densities (or limits, in many cases) of other intermediate and high-ions (i.e.\ \siiv, \nv) and their kinematics.  For this analysis we primarily focus on the column density ratios of \ion{N}{5}/\ion{O}{6} for 38/39 \ion{O}{6} components detected in COS spectra that also cover the \nv~ doublet at $\lambda\lambda$ 1238.8, 1242.8 \AA. For all but three of these components,  the column density ratios are upper limits due to the non-detection of \nv, and should be interpreted as such. 

It is entirely possible that \ion{N}{5} does not trace the same gas phase as \ion{O}{6}. The two ionization potential energies, $\sim$ 98 eV and $\sim$ 138 eV, respectively, differ by 40 eV.  For example, one of the IGM absorbers analyzed by \cite{tumlinson05} along the QSO line of sight to PG 1211$+$143 shows broad \ion{O}{6} consistent with collisionally-ionized, warm gas, and weak, but narrow \ion{N}{5}, consistent with photoionization \citep[see also:][]{savage11}. In contrast, \cite{fox09} in their analysis of \ion{N}{5} detected in damped Ly$\alpha$ systems find a detection rate of \ion{N}{5} to be only 13\%, and to have roughly equal contributions from broad and narrow components. For the \ion{N}{5} components in our sample, all have $b$ values $\approx$ 25 km s$^{-1}$. However, we caution that the  \ion{N}{5} is fairly weak when detected, and the S/N does not allow for very robust profile fits in the few cases of detections.  Of the three \ion{O}{6} components that match up with \ion{N}{5} components, one is in a `narrow' kinematic type absorber, with $b_{\rm OVI}$ = 23 km s$^{-1}$. The other two are found in `broad type' \ion{O}{6} absorbers with $b_{\rm OVI}$ = 72 and 79 km s$^{-1}$.  

For cross comparison with the \ion{O}{6} components, we give their QSO and galaxy identifiers, along with their velocity centroids, log $\nnv$/$\novi$, and kinematic subtype: \begin{enumerate}
\item{J1016$+$4706\_359\_16: $v_{\rm cen}$= $-68$ km s$^{-1}$; \\ log $\nnv$/$\novi$ = $-0.78$\footnote{This absorber has no spectral coverage of \ion{Si}{4} so it is not included in Figure \ref{fig:ratios}.  }, `broad' type \ion{O}{6}}
\item{J1241$+$5721\_199\_6: $v_{\rm cen}$= $65$ km s$^{-1}$; \\ log $\nnv$/$\novi$ = $-1.28$, `broad' type \ion{O}{6}}
\item{J1241$+$5721\_208\_27: $v_{\rm cen}$= $33$ km s$^{-1}$; \\ log $\nnv$/$\novi$ = $-0.83$, `narrow' type \ion{O}{6}}
\end{enumerate}

We suggest inspecting Figure \ref{fig:highprof} on the corresponding panels at the listed velocities. It is not evident by the kinematics that \ion{N}{5}, when detected, traces the same gas-phase as \ion{O}{6}. The component structure of \ion{N}{5} in some cases appears to match that of the low-ionization state gas, and may be more highly structured than the \ion{O}{6}. The S/N of the data make it difficult to draw a more robust conclusion from examining the few detections of \ion{N}{5}. We pointed out in Figure \ref{fig:models} that single-density photoionization models based on low-ions under-predict the column density of \ion{Si}{4}; the same can be said for the three cases of detected \ion{N}{5} absorption.  We do note that the kinematics of intermediate ions such as \ion{C}{4} and \ion{Si}{4} often bear a higher degree of similarity to low-ions than to high-ions \citep[e.g.][]{burchett15}. It is not a given that \ion{O}{6} would exist in this same gas phase as intermediate ions, especially in the cases of `broad' and `no-low' types. Indeed, none of the `no-low' absorbers have intermediate ion detections.  We remain agnostic for now, but must proceed with caution. 

 \ion{N}{5} is not detected in {\emph{any}} gas phase in 35/38 components considered. Thus, its upper limit can be used in comparison against \ion{O}{6} in {\emph{both}} photoionization and collisional ionization models for a single-phase gas. Indeed, we will see that the constraints imposed by the non-detection of \ion{N}{5} in both photoionized gas and collisionally-ionized gas are meaningful. Furthermore, there are two models that attempt to explain {\emph{all}} ion detections in a self-consistent fashion, namely photoionization models that incorporate steep density gradients in the absorbing gas \citep[e.g.][]{stern16}. In those cases, a comparison of the column densities of  \siiv, \nv, and \ovi~is instructive. 

\cite{savage14} have recently reviewed some of the processes other than photoionization that produce \ovi, which include: CIE and non-CIE non-dynamical radiative cooling models \citep{edgar86, gs07}, dynamical radiative cooling models \citep{shapiro91, wakker12}, shock ionization models \citep{dopita96}, conductive interfaces between cool clouds and a hot medium \citep{borkowski90},  and non-equilibrium turbulent mixing layers \citep{begelman90, slavin93, kwak10, kwak11}. Many of these  processes have been invoked for the highly ionized component associated to Galactic HVCs \citep{savage00,fox04}. Additionally,  it is possible to produce \ovi\ via photoionization, with the gas very highly ionized by the EUVB, or by radiation from the galaxy in addition to the EUVB that includes soft x-rays \citep{cantalupo10, vasiliev13}. 
For photoionization in equilibrium we use the Cloudy software package (v13.03; Ferland et al. 2013\nocite{ferland13}) and for
non-equilibrium photoionization modeling and collisional ionization calculations, both in and out of equilibrium, we use the tables of \cite{oppenheimer13}.


\subsection{Photoionization by the Extragalactic UV-Background}
\label{sec:photo}

We now consider the photoionization of an infinite homogenous slab of optically-thin gas
illuminated by an incident extragalactic ultraviolet background (EUVB) radiation field.  This type of photoionization modeling successfully reproduces the observations of the low-ionization state ions (i.e. the \ion{Si}{3}-bearing gas; see Werk et al. 2014), and we explore here whether this simple model can be applied to the more highly ionized gas, though at lower gas densities. 

\begin{figure}[t]
\begin{centering}
\includegraphics[width=0.85\linewidth]{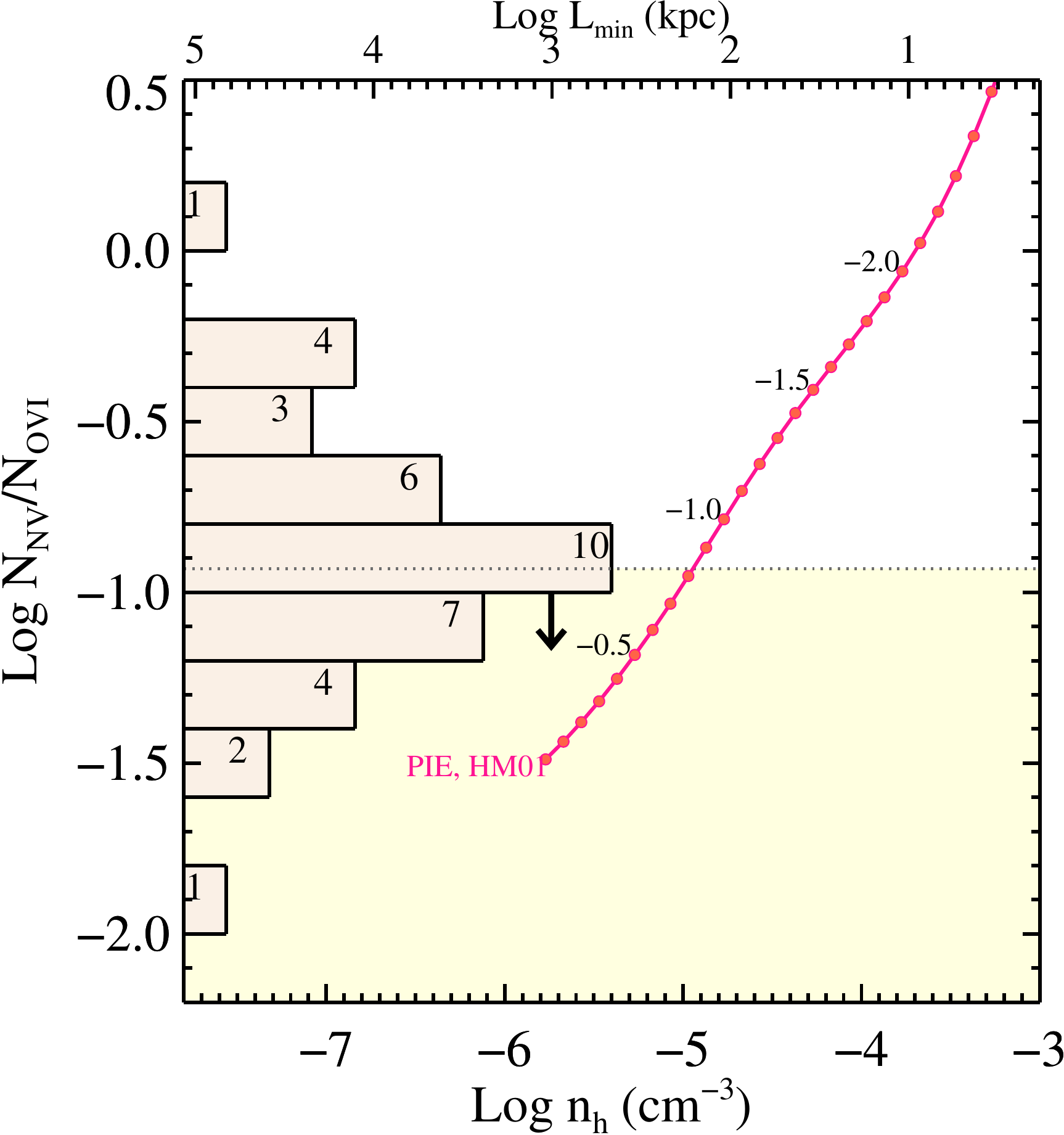}
\end{centering}
\caption{ Column density ratios \ion{N}{5} / \ion{O}{6} as a function of hydrogen volume density for a simple Cloudy-based model shown in pink in which a single-density slab of Z = Z$_{\odot}$, and log N$_{\rm HI}$ = 15 cm$^{-2}$ gas is photoionized by a HM2001 EUVB. The corresponding minimum path lengths are shown on the top x-axis.  We show a histogram of the observed ratios for the individual components on the y-axis for reference, noting with the downward-facing arrow that the vast majority of these column density ratios are upper limits due to the non-detection of \ion{N}{5} for the majority of the \ion{O}{6} components. The pink model curve traces a series of points that correspond to the  dimensionless gas ionization parameter, log U, and we label several values along the curve. The yellow-shaded area highlights the typical COS-Halos measured column density ratios of \ion{N}{5} / \ion{O}{6} $<$ $-$0.93.  Gas ionization parameters with log U $>$ -0.8 are allowed by the majority of the data.     }
\label{fig:simplephoto}
\end{figure}

\begin{figure*}[t!]
\begin{centering}
\hspace{0.5in}
\includegraphics[width=0.90\linewidth]{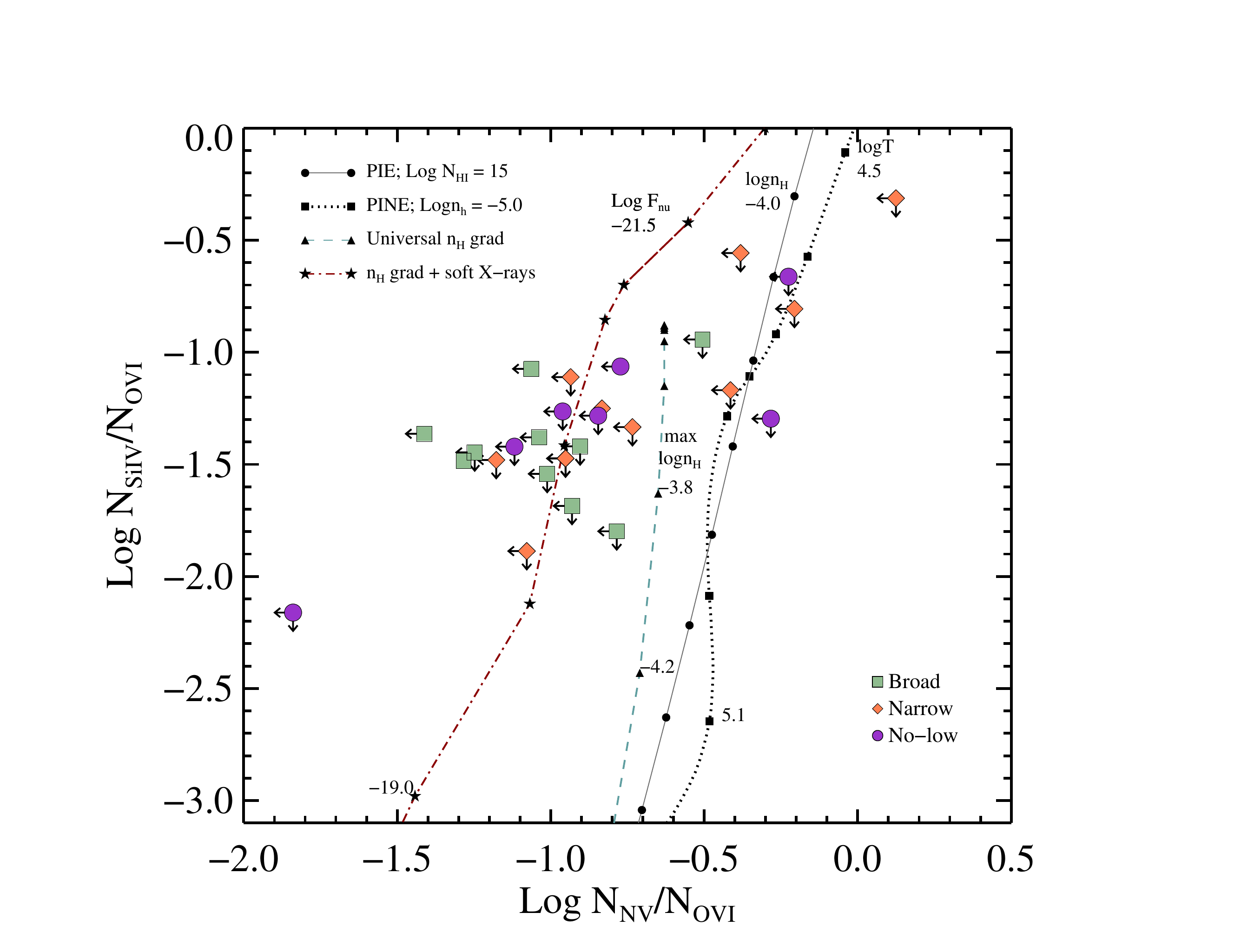}
\end{centering}
\caption{ Ionic ratios, log $ \nnv/\novi$~versus log $\nsiiv/\novi$, for the matched components. As in previous figures, green squares indicate `broad' type \ion{O}{6} absorption; orange diamonds show `narrow' type absorption; and purple circles mark the `no-low' type absorption. We show the predictions of various photoionization models as curved lines, for reference.  The light gray line with black circles on the upper right shows a single-density `slab'  photoionization model in thermal and ionization equilibrium (PIE), photoionized by an HM2001 EUVB, with log U ranging from $-$1.5 to $-$0.5 (log n$_{\rm H}$ $<$ $-4.0$ cm$^{-3}$),  and log N$_{\rm HI}$ = 15 cm$^{-2}$.  PIE models with log N$_{\rm HI}$ $<$ 17 cm$^{-2}$ lie to the right of the line shown, and are therefore less consistent with the data.  We show a model of  metal-enriched gas (Z =  Z$_{\odot}$) photoionized by the EUVB cooling out of equilibrium (PINE) for the isobaric case of n$_{\rm H}$T $=$ 50 given by  \cite{oppenheimer13} as the dotted line curve marked by black squares. Isothermal models for lower metallicity and higher gas densities tend to lie to the right of the plotted curve.  The filled squares on the PINE curve represent different temperatures ranging from 10$^{4.5}$ $<$ T $<$ 10$^{5.1}$, with the lowest temperature at the upper right. The red dashed-dot curve with black stars is a model developed by Cantalupo et al. (in prep) that assumes clouds with a steep density gradient that are photoionized by the EUVB and an extra blackbody component to the ionizing radiation with T$=$ 10$^{5}$ K. The stars on this curve correspond to different intensities of the extra component ranging from log 4$\pi$J$_{\nu}$ at 10 Ryd of  $-$22.0 (top right) to $-$19 ergs s$^{-1}$ cm$^{-2}$ Hz$^{-1}$ (lower left). Finally, the blue dashed line with triangles shows the predictions of \cite{stern16} for the phenomenological universal density model in which small high-density clouds are hierarchically embedded in larger low-density clouds. The triangles mark values of the maximum log n$_{\rm H}$, which occurs in the cloud core. Where log n$_{\rm H, max}$ $>$ -3.0,  the line ratios saturate. 
}
\label{fig:ratios}
\end{figure*}

Figure \ref{fig:simplephoto} shows log $\nnv$/$\novi$ as a function of hydrogen volume density for a simple Cloudy-based model with the EUVB radiation field of \cite{hm01}. The pink line tracks a fiducial model of optically thin Z = Z$_{\odot}$ gas where log N$_{\rm HI}$ = 15 cm$^{-2}$, since most of the HI seen in COS-Halos is likely associated with the low-ions. Along the y-axis we show the distribution of log $\nnv$/$\novi$ for the 38 components, 35 of which are upper limits.  The histogram is labeled according to the number of components in each bin. We note that all three kinematic subtypes show very similar distributions of their upper limits to log $\nnv$/$\novi$.  Therefore it is not instructive to color the histogram by the \ion{O}{6} kinematic component type (though Figure \ref{fig:ratios} shows the actual ratios colored by kinematic type for reference). 

The EUVB we use has a total photon flux $\Phi_{\rm HM01} = 4.8 \times 10^{4}~\rm  cm^{-2}~s^{-1}$  \citep{hm01} for a series of assumed dimensionless ionization parameters $\log U \equiv \Phi/n_{\rm H} c$, which are shown and labeled along the pink line.  The range of log $\nnv$/$\novi$ allowed by the majority of the data is shaded light yellow.  The observations require $\log$ U $\ga$ $-1$ to achieve crude consistency with the data.   Such high $\log \rm U$ values are
characteristic of gas close to quasars and would require a very low gas density for our COS-Halos systems. 
Regarding $\Phi$, the average EUVB is well bounded to have log $\Gamma$$_{HI}$ $\sim
13$ \citep{dave99,davetripp01} and cannot conceivably provide a high U value on
its own, i.e.\ for gas with a modest density, $n_{\rm H} > 10^{-5} \cm{-3}$.

Figure \ref{fig:simplephoto} also shows a minimum  path length for this highly-ionized gas on the top y-axis, computed as N$_{\rm O}$/n$_{\rm O}$, and given in kpc. For example,  a value of $\log U > -1.0$ would require n$_{\rm H} \lesssim 10^{-5} \cm{-3}$ or $n_{\rm O} < 10^{-8.3} \cm{-3}$, 
for a solar metallicity gas.  This gas volume density implies a length scale L $\sim
\novi/n_{\rm O} > 20$ kpc based on $\novi$ alone. We conservatively convert $\novi$ to N$_{\rm O}$ assuming the maximum fraction of gas in \ion{O}{6} allowed by the photoionization models (or any model, because of the cooling function), 0.25, implying an additional factor of $\sim 4$ in size. Gas metallicities below solar would increase this minimum length scale, to e.g. 1 Mpc for Z/Z$_{\odot}$ = 0.1.  This conservative length-scale thus likely exceeds the extent of the dark matter halos hosting \lstar\ galaxies. Furthermore,  these constant density models would be  severely challenged to yield the observed  coherence in velocities (e.g. Figure~\ref{fig:oxysillyprof}), especially for the `narrow' kinematic types.  We conclude that {\emph{the standard EUVB, constant-density photoionization models are ruled out altogether for the \ovi-bearing gas around L$^*$ star-forming galaxies.}}

  Figure~\ref{fig:ratios} presents the observed ionic ratios \siovi\ versus \nvovi\ for 28 individual matched \ion{O}{6} components along sightlines that cover both \ion{N}{5} and \ion{Si}{4}. The data points are colored by kinematic type, though each kinematic type traces roughly the same range of \nvovi\ and \siovi\ especially given that the majority of these points are upper limits on both axes.  Non-detections in \ion{N}{5} give a characteristic upper limit (left-facing arrows) $\log(\nnv/\novi) \lesssim$ -0.9, set by the S/N of the COS spectra and the measured $\novi$ values. The ratio of \siovi\ tends to lie around $-$1 or lower, in the cases of upper limits. We note that the majority of the data points shown in  Figure~\ref{fig:ratios} should occupy a region of parameter space to the left of and/or below the locus of points. The lines on Figure~\ref{fig:ratios} present the column density ratios predicted by several photoionization models that we will explore in detail.  For reference, one of these models (light gray line, with black circles) is the simple Cloudy photoionization equilibrium model already ruled out, above.

We use the models for photoionization out of equilibrium (PINe) of \cite{oppenheimer13}, which tracks the ionization states of a suite of the atoms most important for gas cooling and calculates the non-equilibrium cooling rates. They assume equilibrium initial conditions (t = 0) of n$_{\rm H}$ = 10$^{-4}$ cm$^{-3}$ and T = 10$^{6}$ K. Within 1 Gyr the gas temperature drops below T $<$ 10$^{5}$ K.  In the presence of efficient cooling, for the metal-enriched gases of the CGM (assumed to be 1/10 Z$_{\odot}$), the observational diagnostics are altered somewhat from the equilibrium case. In Figure \ref{fig:ratios}, the dotted black line shows the isobaric PINe model of \cite{oppenheimer13} at n$_{\rm H}$T = 10 K cm$^{-3}$. The other PINe models (including the isochoric models) fall to the right of this curve. PINe alleviates some of the concern of the equilibrium case (PIE, gray line), in that the density does not need to be extremely low to match the data points with the highest allowable \nv/\ovi. In isobaric PINe, the density increases when cooling at a constant pressure, which allows for lower U. Yet, PINe cannot explain the majority of the observed ratios (or limits), because like PIE,  it over-predicts the ratio \ion{N}{5}/\ion{O}{6}. 

\subsection{Photoionization Models Including a Density Gradient in the Absorbing Cloud}

Let us now consider a departure from typical photoionization models that assume a homogenous slab of gas with a constant density, and instead allow a density gradient that increases toward the center of the absorber. The low-ions will be located in the central compact, denser regions, while the high-ions will lie in the outer, lower-density gas.  The specific location of the line of sight intercepting the cloud will affect the observed properties of the absorber.

 \cite{stern16} explores such a phenomenological model in detail, presenting a universal density gradient (UDG) for cool clouds in the CGM that explains the gas column densities of the full suite of ions covered by COS-Halos, from \ion{Mg}{2} to \ion{O}{6}. We show the expectation of log $\nsiiv/\novi$ vs. log $\nnv/\novi$  for their model in Figure \ref{fig:ratios} as the blue dashed curve marked with triangles for values of the maximum log n$_{\rm H}$, or the maximum density of the absorber along the line of sight. The typical length scales and densities predicted by the model have a range of approximately three orders of magnitude from the inner `cloud' to the outer low-density outskirts. For example, the UDG model predicts \ion{O}{6} lives in the outermost layer of the cloud, characterized by a gas density of $\sim$2 $\times$ 10$^{-5}$ cm$^{-3}$, and path lengths $\sim$30 kpc. In contrast, \ion{Mg}{2} is confined to the inner 49 pc and has a density of  $\sim$6.5 $\times$ 10$^{-3}$ cm$^{-3}$ \citep{stern16}. 

Figure \ref{fig:ratios} shows that $>$ 50\% of the upper limits to the log $\nnv/\novi$ are inconsistent with the predictions of the UDG. The mean {\emph{upper limit}} to  log $\nnv/\novi$ for the 38 components that cover \ion{N}{5} is $-0.9$, while the value predicted by the universal density model is typically $\sim-0.8$ for a large range of maximum gas densities. Figure 4 of Stern et al. shows that the UDG slightly over-predicts upper limits to N$_{\rm NV}$, which is the source of the discrepancy. As we noted in Section \ref{sec:nitrogen},  the single-density photoionization models of COS-Halos data are broadly consistent with solar N/O ratios. However, it may be possible to further tune the UDG to include sub-solar N/O ratios (and still account for the observed \ion{N}{2} and \ion{N}{3}). Such an adjustment might be one avenue by which to bring the UDG into agreement with the constraints placed on log $\nnv/\novi$ by the COS-Halos data.

One appealing facet of the UDG model is that it simultaneously explains the column densities of both low and intermediate ions very well, resolving the tension between \ion{Si}{4} and and the low-ions that is present in single-density Cloudy models. Furthermore, the UDG implied length scales for \ion{Si}{4} are a modest $\sim$4 kpc, which would not necessitate large differences in the velocity structure of the \ion{Si}{4} absorption compared to the \ion{Si}{3} absorption (see Figure \ref{fig:highprof}). Finally, the UDG brings the column density predictions for \ion{Si}{3} and \ion{Si}{2} into better agreement than found in single-density models. However, density gradients may be difficult to maintain for timescales greater than the sound-crossing time ($\sim$ 10$^{6}$ years for 10$^{4}$ K gas) without continual shocking. Future studies addressing the physical origin of the UDG will have to discuss the formation and maintenance of the required gradient, without significant contribution from collisional ionization in the wakes of shocked material.

\subsection{Additional Sources of Radiation}
\label{sec:xray}
\label{sec:extraphot} 
 \label{sec:cantalbb}
 
Metal-line diagnostics, such as the ratio of \nv~to \ovi, depend critically upon the spectral shape and strength of the ionizing radiation. The EUVB is calculated assuming galaxy escape fractions, intrinsic spectral slopes of QSOs, and extrapolations of luminosity functions, all of which are uncertain \citep{hm01, hm12}. For example, at the energy required to ionize \ion{O}{5} to \ion{O}{6} (114 eV), the uncertainty in the shape and intensity of the EUVB can contribute up to an order of magnitude uncertainty in the derived ionic abundances \citep{oppenheimer13}. This is true also at the energy where \ion{N}{4} is ionized to \ion{N}{5} (77 eV). By extension, the uncertainty of the contribution of ionizing radiation from the host galaxy can have a major impact on the metallic ion ratios. Moreover, hot gas is capable of producing significant additional  `self-ionizing' radiation \citep{gnat10}, which in turn depends on the mechanism of ionization in the hot gas. Here we explore the effect of adding additional sources of ionizing photons to the EUVB, which will change both the overall intensity and shape of the incident radiation field. 

 As the absorption we observe occurs within 160 kpc of a nearby galaxy, the EUVB may be supplemented or even exceeded by a local ionizing radiation field from sources within the galaxy. The simplest implementation of this idea is to scale the total ionizing flux, $\Phi_{\rm tot}$, with impact parameter ($\propto$ 1/R$^{2}$), galaxy star formation rate ($\propto$ SFR), and the escape fraction of ionizing photons ( $\propto$ f$_{\rm esc}$).  Moreover,  there is some observational motivation for considering such a local ionizing radiation field given the correlation observed for `broad ' type absorbers between N$_{\rm OVI}$ and SFR/R$^{2}$, as mentioned in Section \ref{sec:galprop}. Such an increase in $\Phi_{\rm tot}$ would ease the requirements for low gas density found in \cite{werk14}, and the implied huge length scales discussed in Section \ref{sec:photo}  because n$_{\rm H}$ $\propto$ $\Phi_{\rm tot}$/U. Yet, the inclusion of only stellar radiation from a Starburst99 spectrum \citep{lsg99} will not provide the increase in $\log$ n$_{\rm H}$ required unless the SFRs are extreme. For example, if the galaxy SFR exceeds 50 M$_{\odot}$ yr$^{-1}$ for sightlines with  R$<$ 50 kpc  assuming f$_{\rm esc}$ $=$ 5\%, the gas density will increase by two orders of magnitude. For reference, the average SFR for the COS-Halos star-forming galaxies is approximately 1 M$_{\odot}$ yr $^{-1}$ and the average impact parameter is 72 kpc. We must therefore consider additional sources of ionizing radiation that are not included in the starburst99  galaxy SED (Leitherer et al. 1999\nocite{lsg99}).  

In their investigation of how photoionization by local sources regulates gas cooling, \cite{cantalupo10} use a galaxy SED that  incorporates soft X-ray emission produced by mechanical energy released into the ISM during a starburst phase, from both supernovae and additional X-ray sources produced by star-formation. This SED is calibrated to reproduce the empirical relation between soft X-ray emission and SFR presented by \cite{ceverino02}. Along with other ionizing radiation from the galaxy, some fraction of the soft X-rays escape into the CGM. Though less physically-motived, the SED of \cite{vasiliev13} also shows an extra soft X-ray component.  In this case, the authors simply sum together a PEGASE SED, an X-ray power-law spectrum, and an EUVB in a time-dependent radiation field to investigate how a local radiation field including X-rays impacts the ionization fraction of \ion{O}{6}.  The break in their spectrum at 91 \AA~ is due to the lack of data at higher energies in the PEGASE data rather than having a physical origin. Nonetheless, both the  \cite{cantalupo10} and \cite{vasiliev13} SEDs, when added to the EUVB, produce an excess of high-energy photons ($>$ 8 Ryd) relative to those with energies between  4$-$8 Ryd. The latter are partially absorbed by the galaxy ISM, with He and dust being important sources of opacity. 
  
 In Figure \ref{fig:ratios},  we show a toy model that includes an excess of radiation at energies $>$ 50 eV  as the red dashed-dot line with stars.  For simplicity,  we make the following assumptions (a more detailed model exploring a larger parameter range will be presented in Cantalupo et al., in prep.):  (1) The SED of the ionizing radiation is a composite of a blackbody with  T$=$10$^5$ K,  absorbed intrinsically by a galaxy with N$_{\rm HI}$=10$^{20}$ cm$^{-2}$, and an EUVB at z$=$0.2; (2) The cloud is a plane parallel slab with total column of N$_{\rm H}$=10$^{20}$ cm$^{-2}$, with an inwardly-increasing density distribution having an initial volume density, log n$_{\rm H_{\rm 0}} = -3.5$ with the scale factor N$_{\rm H_{\rm 0}}$=10$^{19}$. This density profile obeys the relation: n(r) $=$ n$_{\rm H_{\rm 0}}(r_{\rm 0})\times $ (1 +  N$_{\rm H}$/N$_{\rm H_{\rm 0}}$)$^{2}$ [cm$^{-3}$] and is less steep than the universal density gradient explored by \cite{stern16}. The density decreases by a factor of $\sim$100 toward the outer regions the cloud. The model, shown as the red dashed-dotted line in Figure \ref{fig:ratios}, does well in reproducing the  ionic ratios \siovi\ and \nvovi\  and the total column densities for the majority of the absorbers\footnote{This model matches the low ionization state lines well, since the SED below $\sim$ 10 Rydberg is unchanged.}.  Each star represents a unique value of the intensity of the extra blackbody SED component with values of log 4$\pi$J$_{\nu}$ at 10 Ryd between $-$22.0 (top right) and $-$19 ergs s$^{-1}$ cm$^{-2}$ Hz$^{-1}$ (lower left). The predictions of  $\log \novi$ range from 13.52 to 15.1, coincident with the observed values. 

The main utility for this toy model lies in constraining the intensity of the extra component in conjunction with the gas density (ionization parameter) required for photoionization to remain a viable model for the COS-Halos \ion{O}{6} absorbers:  $\log 4\pi$J$_{\nu}$ at 10 Ryd $>$ $-$21.5 ergs s$^{-1}$ cm$^{-2}$ Hz$^{-1}$ for log n$_{\rm H}$ = $-3.5$ cm$^{-3}$.  To first order, the required J$_{\nu}$ will scale with density. For reference, the EUVB produced by quasars and galaxies at z$\sim$0.2 provides log 4$\pi$J$_{\nu}$ at 10 Ryd $=$ -23.7 for HM01 and -23.2 for HM12 \citep{hm01, hm12}. As we have seen, the EUVB is by itself orders of magnitude too weak to account for the observed ionization states of the CGM gas if photoionization is the dominant mechanism. The success of the Cantalupo model relies also on the density gradient, since \ion{Si}{4} is partially shielded by \ion{H}{1} and tends to live in the denser part of the cloud with the other low-ions. 
 
This model further necessitates that \ion{O}{6} reside in a cool, $<$10$^{5}$ K phase. The large path lengths implied by EUVB-only photoionization ($\gtrsim$ 100 kpc) models are now mitigated by the increase in $\Phi$, and on the order of $\sim$ 1$-$10 kpc (see Section \ref{sec:photo}).  In the COS-Halos dataset,  \ion{O}{6} is often significantly broader than the low-ions, possibly indicating a temperature limit that is $>$ 6 times as high as that for the low-ions.  However, if large scale motions of the gas or turbulence contribute significantly to the line width, then these assumptions about temperature inferred from the line widths are incorrect. In general, for `narrow' and `broad' \ion{O}{6} kinematic subtypes, the gas kinematics are consistent with \ion{O}{6} arising in the same structure as the low-ions, since the offsets in the velocity centroids are typically $\sim$20 km s$^{-1}$. 

Using  $\log 4\pi$J$_{\nu}$ $\gtrsim$ $-$21.5 ergs s$^{-1}$ cm$^{-2}$ Hz$^{-1}$, we may constrain the luminosity of the unknown sources the model requires at $\sim$10 Ryd and compare with that of potential candidates for the emission.  At distances $>$10 kpc, this luminosity $\nu$ L$_{\nu}$ $>$ 10$^ {40}$ erg s$^{-1}$ at $\sim$10 Ryd. Unfortunately, this emission energy,  $\approx$ 0.13 keV, is outside of the classical bands of X-ray instruments (e.g. Chandra), and our own galaxy typically absorbs most of the extragalactic radiation at these energies. The main sources of radiation at these soft X-ray energies, excluding fluctuating AGN \citep[e.g.][]{oppenheimer13b} and possible contribution from X-ray binaries, are expected to be the hot ISM (heated by supernovae) and the so-called ``supersoft X-ray sources (SSSs)" \citep{greiner00}.  Soft X-ray emission from the ISM is known to correlate  with the galaxy SFR,  and in all star forming galaxies there is at least one optically thin thermal emission component with an average temperature of $<kT>$ = 0.24 keV \citep[e.g.][]{mineo12}.  Although the derived intrinsic bolometric luminosity from this gas is consistent with our model requirements at d $>$ 10 kpc,  the luminosities are model dependent and require extrapolation to the lower energies. Additional sources of high-energy photons could arise from the SSSs. First identified more than two decades ago by the Einstein Observatory, these close binaries are characterized by significant emission in the $0.3-0.7$ keV band. Their SED resembles the tail of a blackbody with effective temperatures of a few $\times$ 10$^5$ K \citep{greiner00}. In the disks of ordinary star forming galaxies like the Milky Way and M31, there are estimated to be of the order $\sim$10$^3$ SSSs, thus providing enough intrinsic emission at 10 Ryd compared to our constraints \citep[e.g.][]{kahabka97}. Taken at face value, our results would imply that most of the radiation produced by SSSs should escape from the ISM \citep[in agreement with the results of][]{woods16}.  As explored by \cite{cantalupo10}, the presence of such an intense radiation field at high energies has a profound impact on the cooling of the CGM and therefore on galaxy formation and evolution as we discuss in Section \ref{sec:discussion}. 


\subsection{Collisional Ionization} 
\label{sec:cie}

Now we consider a non-dynamical gas which radiatively cools via collisions both in and out of equilibrium (CIE and CINe). These models are parameterized primarily by temperature. In the equilibrium case,  the cooling efficiency is dependent upon the gas temperature, density, and composition. In the absence of an additional heat source, the gas cools via the removal of electron kinetic energy due to recombinations with ions, collisional ionizations and excitations (followed shortly by line emission), and thermal bremsstrahlung.  For non-equilibrium cooling, the gas is``overionized" compared to CIE because cooling is more rapid owing to a ``recombination lag" \citep{gs07}.
 
 
 \begin{figure}[t]
\begin{centering}
\hspace{0.0in}
\includegraphics[width=0.85\linewidth]{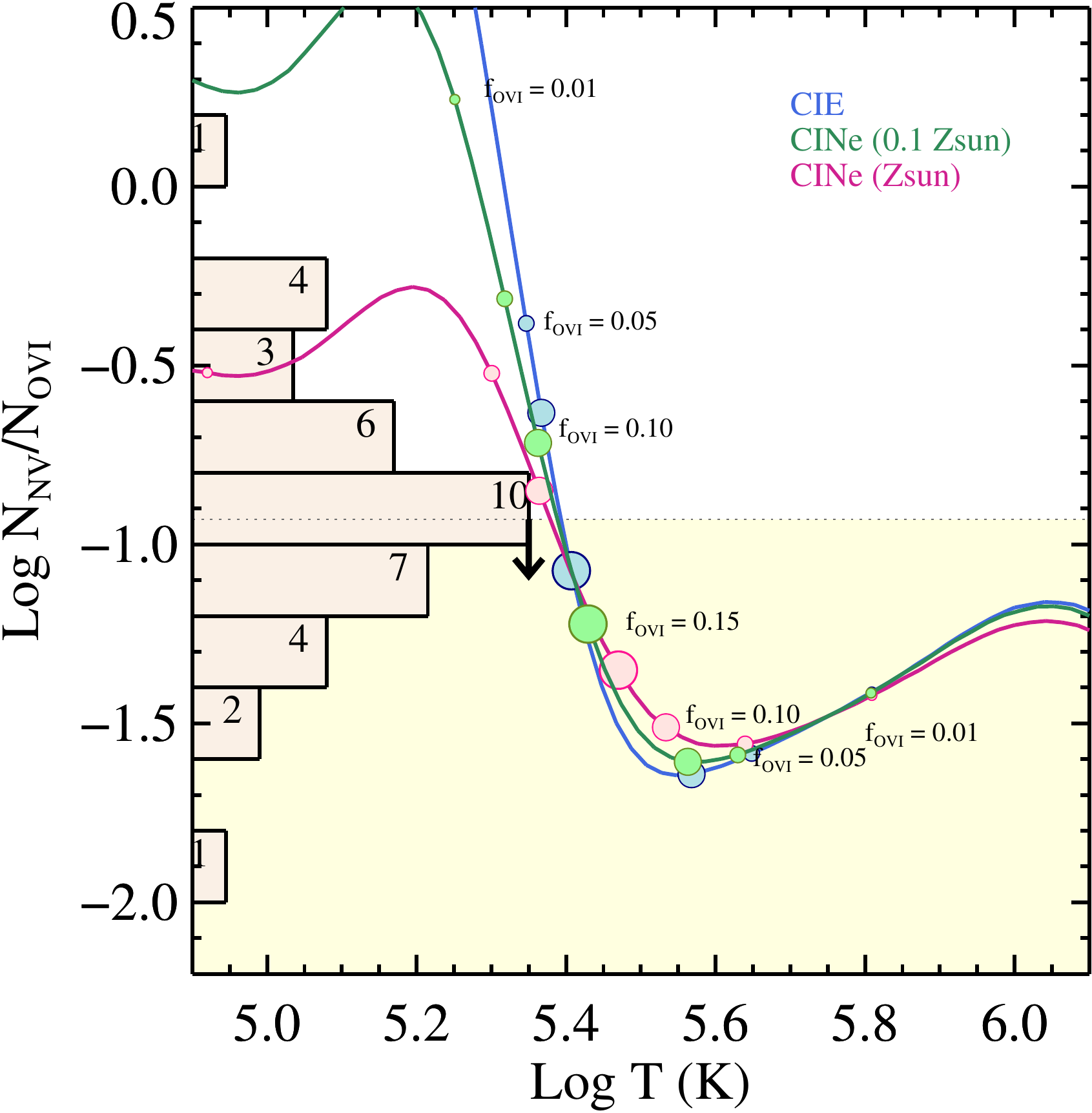}
\end{centering}
\caption{ log N$_{\rm NV}$/ N$_{\rm OVI}$ as a function of gas temperature for collisionally-ionized gas both in (CIE) and out of equilibrium (CINE). The curves shown represent the models of \cite{oppenheimer13}, and the filled circles on each curve give the fraction of oxygen in the \ion{O}{6} state which varies with temperature. f$_{\rm OVI}$ is maximized close to the minimum allowed value of log N$_{\rm NV}$/ N$_{\rm OVI}$ $\approx$ -1.5.  We show a histogram of the observed ratios for the individual components on the y-axis for reference, noting with the downward-facing arrow that the vast majority of these column density ratios are upper limits due to the non-detection of \ion{N}{5} for the majority of the \ion{O}{6} components. The yellow-shaded area highlights the typical COS-Halos measured column density ratios of \ion{N}{5} / \ion{O}{6} $<$ $-$0.93. CIE and CINE models indicate that for the observed  limits on log N$_{\rm NV}$/ N$_{\rm OVI}$, gas temperature lies within the range 5.4 $<$ log T $<$ 5.6 and  the \ion{O}{6} gas fraction falls within 0.10 $<$ f$_{\rm OVI}$ $<$ 0.15. }
\label{fig:collisional}
\end{figure}

Figure \ref{fig:collisional} shows log $\nnv$/$\novi$ as a function of the gas temperature for collisionally-ionized gas both in (CIE; blue curve) and out of equilibrium (CINE; green and pink curves).  On each curve, we supply the value of the fraction of oxygen in \ion{O}{6} predicted by the model. The observed \ion{N}{5}/\ion{O}{6} ratios, represented on the y-axis as a histogram, require $T > 10^{5.35}$K, which is approximately the temperature where the fraction of oxygen in \ovi\ is maximal. One facet of the CIE and CINe models is that for T $>$ 10$^{5.7}$ K, the fraction of oxygen in \ovi\ drops so low that to reproduce the observed \ovi\ column densities, one requires very high values of N$_{\rm O}$ that would imply metallicities above a few times solar. Thus, because of the shape of the cooling curve, CIE and CINe models require a very narrow range of temperature to reproduce the observations.  The upper limits to the \ion{N}{5}/\ion{O}{6} ratios are broadly consistent with this temperature range, and therefore are reproduced by a variety of collisional ionization models. 

\subsection{Complex Non-Equilibrium Models}

Finally, we consider other ionization mechanisms to reproduce the observed  \ion{O}{6}.
These models are primarily a series of collisional ionization processes
out of equilibrium.  They were introduced first in the context of
highly ionized gas gas detected within the Galactic halo, including
gas associated with HVCs \citep{shapiro76}. Following previous work, we consider the predictions for
a wide range of parameter space given by each model. 

\subsubsection{Radiative Cooling Flow}

As shown in Figure \ref{fig:Nb}, a radiative cooling flow model can give rise to a correlation between $\novi$~and its Doppler $b$ parameter under the assumption that the $b$-value reflects a flow velocity  \citep{edgar86}. \cite{wakker12} describe in detail the model first proposed by \cite{shapiroandmoore76} and later developed by \cite{edgar86, shapiro91} and \cite{benjamin94}.  This model allows gas to cool via collisions as it moves at a constant speed.  For a range of flow velocities (v$_{\rm flow}$ $<$ 30 km s$^{-1}$), the model tends to give narrow ranges of log $\nvovi$ between $-1.25$ and $-1.0$, and $\log \siovi$ of $-1.9$ to $-0.2$ (see also Wakker et al. 2012 for a discussion of this model). We note that nearly all of the observed \ion{Si}{4} is generated by photoionization from the warmer, upstream \ion{O}{6}-producing gas. Such self-photoionizing radiation may be problematic in a cosmological context, however, which we discuss in Section \ref{sec:discussion}. Furthermore, this model does not include radiation from the EUVB. The gas bearing  \ion{N}{5} and  \ion{O}{6} is cooling via collisional ionization at T $\approx$ 10$^{5.5}$ K. 

While the ratios predicted by the radiative cooling flow model are remarkably consistent with the COS-Halos intermediate and high-ion data, the results remain inconclusive with respect to $\novi$ and line-of-sight kinematics. The one-dimensional cooling flow model of \cite{shapiro91} was calculated solely in the case of observing the gas from the Milky-Way, and thus only along the direction of the flow. Although log $\novi$ is 14.5 in this geometry, consistent with the data, it is clear that the angle between the line of sight and the flow direction will impact the predicted column densities, and drive them down considerably. 

\subsubsection{Turbulent Mixing Layers}

First proposed by \cite{begelman90}, later developed by \cite{slavin93}, and expanded upon by \cite{kwak10}, the turbulent mixing layer model involves cool clouds moving through a hot medium possess a shell of gas at their boundary in which Kelvin-Helmholtz instabilities are mixing the hot and cool gas \footnote{\label{note1}This relation can also give rise to an N-b relationship}. This turbulent mixing produces gas in the shell with an intermediate temperature, characterized by highly ionized species. This model is referred to as a turbulent mixing layer (TML), and is commonly invoked to explain the ionization state of  Milky Way HVCs \citep{sembach03, fox04, wakker12, savage14}, and intergalactic intervening absorption \citep{tripp08}. Here, we consider the predicted column density ratios given by \cite{slavin93, kwak10}, which are generally a function of two variables:  the turbulent velocity, v$_{\rm turb}$, which  is allowed to range from 25 $-$ 100 km s$^{-1}$,  and the mixing layer temperature, T$_{\rm mix}$, which ranges from 10$^{5.0}$ $-$ 10$^{5.5}$ K.  The update by \cite{kwak10} includes a more detailed treatment of non-equilibrium ionization in a two-dimensional hydrodynamical model which has a small effect on the ratios, and a more significant effect on the total column densities.  The resultant \nvovi\ ranges from $-1$ $-$ $-0.4$. However, the typical $\novi$~in \cite{slavin93} is $\sim$10$^{12}$ cm$^{-2}$, requiring more than 300 such interfaces to exist along the line of sight to match the observed $\novi$~($\sim$10$^{14.5}$ cm$^{-2}$). In the \cite{kwak10} model, the average line of sight $\novi$~increases to 10$^{12.8}$ cm$^{-2}$, easing the requirement for the number of clouds along each line to sight to be  $\sim$50 to match the COS-Halos \ion{O}{6} column densities. Nonetheless, this physical picture is difficult to reconcile with the observed absorption profiles, which show an average of 2.4 components per line of sight over a fairly narrow range of velocity. 

\subsubsection{Conductive Interfaces}

A cool cloud embedded in a hot medium can also produce a surface layer in which cool gas is evaporating and hot gas is condensing because electron collisions are conducting heat between the two media \citep{borkowski90, gnat10}. Referred to as conductive interfaces, these models predict how the column densities of transition temperature ions change as a function of time and the angle between the magnetic field and the conduction layer orientation. Along magnetic field lines, the thermal conductivity of the interface is much greater, increasing the column densities of the high-ions. The thermal conduction at the interface layer may prevent formation of Kelvin-Helmholtz instabilities at the cloud-corona interface, ultimately leading to long cloud survival times \citep{armillotta16}. Before  2 $\times$ 10$^{5}$ years, the high-ion column densities change rapidly, and then stay relatively constant for the next  $\sim$ 5 Myrs. As \cite{wakker12} points out, this rapid evolution results in a unique value of the predicted ion ratios for 90\% of the lifetime of the interface. The updated conductive interface models of \cite{gnat10} include a range of temperature and pressure of the surrounding hot medium of 10$^{6}$ $-$ 10$^{7}$K, and 0.1 $-$ 50 K cm$^{-3}$, respectively. They also include photoionization by the EUVB. \cite{wakker12} have found that the expanded parameter space of the updated models changes the predicted column densities by an average of 0.2 dex from the original \cite{borkowski90} results. 

Conductive interfaces can have $\log$ \nvovi\ as low as -1.25 for B$_{\parallel}$ and -1.0 for B$_{\perp}$. The ratio $\log$  \siovi\  is -2.0 for B$_{\parallel}$ and -1.2 for B$_{\perp}$. These ratios are consistent with many of the COS-Halos data points. Yet, in these models the column density of  \ion{O}{6} never exceeds 10$^{13}$ cm$^{-2}$ \citep{gnat10}, requiring 20 $-$ 100 interfaces along each line of sight for consistency with the absorption-line data. For different model parameters (i.e. hot medium with T $<$ 10$^{7}$ K), $\novi$ can be as low as 10$^{10}$ cm$^{-2}$, and tens of thousands of interfaces are required to bring the model and data to consistency. Thus, we reach a similar conclusion with respect to conductive interfaces as we did for turbulent mixing layers. There is simply not enough total column density of high ions produced in this model for it to be a plausible explanation for the abundance of  \ion{O}{6} absorption observed in the halos of star-forming galaxies. 

\subsubsection{Supersonic Shocks}

Another way to produce a significant amount of  \ion{O}{6} absorption is to consider collisionally ionized gas behind a high speed radiative shock \citep{dopita96, heckman02, allen08, gnat09, grimes09}.\footnotemark[\ref{note1}] Under the assumption of a steady, one-dimensional flow, and a given metallicity, several models predict column densities of high ions in the post-shock cooling layers. A generic feature of these models is that $\novi$ in the post-shock gas stays below $\sim$10$^{12}$ cm$^{-2}$ for v$_{\rm shock}$ $<$ 150 km s$^{-1}$, and then abruptly increases by two orders of magnitude at v$_{\rm shock}$ $=$ 175 km s$^{-1}$. Here we use the results of \cite{allen08} who produce tables that are valid for solar metallicity, shock speeds that range from 100 $-$ 1000 km s$^{-1}$ and pressure exerted by the transverse magnetic field (Bn$^{-\frac{1}{2}}$) ranging from 10$^{-4}$ $\mu$G cm$^{-3/2}$ - 10 $\mu$G cm$^{-3/2}$. The magnetic pressure is important because it limits the compression through the shock, and thus higher magnetic parameters produce higher $\novi$.  We consider the full model range for the magnetic parameter of the gas.   \cite{gnat09} has updated these models for a wider range of metallicity, though they only consider faster shocks with v$_{\rm shock}$ of 600 and 2000 km s$^{-1}$. 

Generally, $\novi\ $ remains at some level for the first $\sim$ 3 Myr after the initial shock, and then abruptly increases to by two orders of magnitude during the non-equilibrium cooling phase. The total $\novi\ $ in the post-shock gas can range from 10$^{14.3}$ cm$^{-2}$ to 10$^{15.4}$ cm$^{-2}$ for weak magnetic pressure to strong magnetic pressure over the range of velocities considered. \nv\ has a similar broad range, though is consistently 1 $-$ 1.5 dex lower than the total $\novi$, consistent with the COS-Halos data. The kinematics of \ion{O}{6} in this model would likely be varied and complex depending on shock speed, as is observed. And furthermore, if these supersonic shocks arise as a result of starburst-driven winds, this model provides a natural explanation for the correlation between $\novi$ of the `broad' \ion{O}{6} absorbers and the SFR/R$^{2}$.   Thus, collisionally-ionized gas cooling behind a fast shock remains a viable physical model for the observed \ion{O}{6}, and we consider its broader implications for gas kinematics and cooling times in the discussion. 
 
\section{Summary of Results}
\label{sec:summary}

 We have analyzed the \ion{O}{6} kinematics and column densities along 24 lines of sight probing the halos of low-redshift, L$\sim$L$^*$ star-forming galaxies at one-dimensional projected distances of 15 - 150 kpc. We present a simple, data-driven method of identifying three distinct \ion{O}{6} kinematic subtypes present in the {\emph{HST}}/COS FUV spectra, which we enumerate below. Each kinematic subtype shows different behavior with respect to its total column density and host galaxy properties like SFR, R, and M$_{\rm halo}$. For the majority of \ion{O}{6} absorbers in L$^*$ galaxy halos, regardless of their kinematic subtype, photoionization by the EUVB-only is strongly disfavored due to the implied path lengths $>$ 100 kpc. Furthermore, for every kinematic subtype, upper limits to $\nnv/\novi$ place strong constraints on the physical state of the gas under the conditions of photo and colllisional ionization. Finally, we infer that there are multiple distinct physical processes that lead to the observed widespread \ion{O}{6} in the CGM of star-forming galaxies. 
 
The salient observed properties of the three kinematic subtypes are: 
 
\begin{enumerate} 

\item{The `no-low' type absorbers (8/39) are typically broad ($b \approx 50$ km s$^{-1}$) \ion{O}{6} absorption lines with no underlying, detectable absorption from lower ionization state metal species. These 8 absorbers follow a statistically significant, fairly tight correlation between their column density and line-widths, consistent with predictions for radiatively cooling gas in which the total column density is set by a characteristic cooling flow velocity (e.g. gas cooling behind a shock, Dopita \& Sutherland 1996; mixing turbulently in a fixed layer between cool and hot gas, Begelman \& Fabian 1990; collapsing in a cooling instability, Heckman et al. 2002). The upper limits on $\nnv/\novi$ strongly suggest a gas temperature of T $\approx$ 10$^{5.5}$ K under the assumption of collisional ionization. {\emph{These `no-low' type absorbers are found strictly in the COS-Halos galaxies with 10$^{11.5}$ $<$ M$_{\rm halo}$ $<$ 10$^{12}$ M$_{\odot}$, where the halo virial temperature is also T $\approx$ 10$^{5.5}$ K.}} }  

\item{The `broad' type \ion{O}{6} absorbers ($b \gtrsim 40$ km s$^{-1}$) that {\emph{are}} coincident with relatively narrow low-ion absorbers are the most common kinematic subtype found in the CGM of star-forming galaxies (16/39). These `broad' type absorbers alone set the observed trends between $\novi$ and galaxy properties (R, SFR), and dominate the total column density along each line of sight. They are significantly broader than the lower column density \ion{O}{6} lines seen in samples of absorbers observed along blindly selected lines-of-sight (i.e. in the IGM). Their total columns and line-widths are not inconsistent with the radiative cooling relation between log N and $b$, but these absorbers could be impacted by the blending of several narrow components separated by v $<$ 10 km s$^{-1}$. The upper limits on $\nnv/\novi$ strongly rule out most photoionization models, both in and out of equilibrium, both with and without steep density gradients that include radiation from the EUVB only. We use the upper limits of  $\nnv/\novi$ to estimate the total required intensity in addition to that of the EUVB, and find that it must exceed the EUVB by a factor of 100 at energies $>$ 10 Rydberg for photoionization to remain relevant for producing the observed \ion{O}{6}. The upper limits to  $\nnv/\novi$ are consistent with collisional ionization models (in and out of equilibrium), where the fraction of oxygen in \ion{O}{6} is maximized, T $\approx$ 10$^{5.5}$ K.  }

\item{The `narrow' type \ion{O}{6} absorbers ($b \approx 25$ km s$^{-1}$) that are consistent with low-ion absorbers represent 15/39, or 38\%, of the \ion{O}{6} components in the CGM; they are perhaps the most puzzling kinematic subtype. Their total columns and line-widths are {\emph {inconsistent with the radiative cooling correlation}} between log N and $b$, as they systematically lie above it by 0.2 - 0.4 dex.  Their column densities do not correlate with {\emph{any}}  galaxy property observed by COS-Halos. Furthermore, upper limits on $\nnv/\novi$ strongly disfavor an origin in gas photoionized by the EUVB. Although upper limits on $\nnv/\novi$ give T$_{\rm CIE}$ $\approx$ 10$^{5.4}$ K, for collisional ionization models, it is not clear how such models could give the observed tight kinematic correspondence between low and high ionization state gas, i.e. produce \ion{O}{6} line widths generally consistent with those of the low-ionization state material. }

\end{enumerate}


\section{Discussion}
\label{sec:discussion}

Our work shows that even in the relatively well-controlled environment of z$\sim$0.2 star-forming, L$^{*}$ galaxy halos at R $<$ 150 kpc, the highly-ionized \ion{O}{6}-bearing gas appears to have a variety of origins and a range of ionization states. This result may explain why previous works have been unable to come to a generic conclusion about the origin of \ion{O}{6} \citep[but see:][]{heckman02, bordoloi16}. The present study builds on those previous efforts which laid the foundation for the emerging complex picture of highly ionized, diffuse gas in a wide range of environments at a wide range of redshifts \citep[e.g.][]{sembach03, fox04, tumlinson05, fox06, tripp08, fox09, lehner09, wakker09, narayanan10, narayanan11, lehner11, prochaska11, savage11, muzahid12, wakker12, narayanan12,  savage14, lehner14,  hussain15}. Simply, there is no single model that can self-consistently explain the observed variety of \ion{O}{6} kinematic correspondence with low-ionization state gas along with the high-ion absorption-line component column density ratios. 

In this Section, we focus on favored physical and/or phenomenological explanations for each of the distinct kinematic subtypes. We comment on their implications for the total baryonic contribution of \ion{O}{6}-bearing gas to the galaxy halo, discuss gas cooling times, and halo dynamics. 


\subsection{The `No-Low' Absorbers: T $\approx$ T$_{\rm vir}$ Gas}

The theory that there exists a `critical' galaxy halo mass marking a sharp transition in gas cooling rates has persisted since the early days of analytic galaxy formation models \citep{rees77, silk77, binney77}, and been refined by modern cosmological hydrodynamical simulations \citep[e.g.][]{keres05, stinson15, oppenheimer16}. Generally,  galaxies with M$_{\rm halo}$ $<$ M$_{\rm crit}$ are able to accrete cool gas onto their disks, while the accreting gas in galaxies with M$_{\rm halo}$ $>$ M$_{\rm crit}$ shock heats to the virial temperature, and cools over a dynamical time. However, it is now widely recognized by theorists that feedback plays a key role in the regulation of the halo gaseous medium, and the physics of the gas both above and below the critical halo mass threshold \citep[e.g.][]{oppenheimer08, governato10,  fg11, hopkins14, nelson15, muratov15, fielding16, christensen16, oppenheimer16}. 

One of the key discoveries of our kinematic study is that the `no-low' kinematic subtype of \ion{O}{6} absorption abruptly disappears for galaxies with M$_{\rm halo}$ $>$ 10$^{12}$ M$_{\odot}$. This break has a statistical significance of 3$\sigma$ for the 8 `no-low' data points, and is not likely the result of a selection effect.  This sharp mass cut-off appears to be consistent with the critical halo mass below which T$_{\rm vir}$ $\approx$ 10$^{5.5}$ K \citep{oppenheimer16}.  We showed that the no-low \ion{O}{6} absorber is most likely collisionally ionized, by noting its remarkable consistency with radiative cooling predictions (for a wide range of physical scenarios) for an N-$b$ relation, and that its limits on $\nnv/\novi$ are consistent with collisional ionization models for T $\approx$ 10$^{5.5}$ K. We note that the density of this gas must be $\lesssim$ 2 $\times$10$^{-5}$ cm$^{-3}$  for the cooling time to be $>$ 10$^9$ years in CIE models, and hence for gas not to be in dynamical state \citep[calculated at Z = 0.1Z$_{\odot}$; ][]{gs07, oppenheimer13}.  The simplest interpretation of this result is that the `no-low' \ion{O}{6} absorbers are tracing virialized halo gas in the subset of COS-Halos galaxies with M$_{\rm halo}$ $<$ 10$^{12}$ M$_{\odot}$. 

Although the gas is most likely bound to the galaxy halo (See Figure \ref{fig:vesc}),  the large distribution of the `no-low' velocity centroids with respect to the galaxy systemic velocity ($\Delta$v) is not obviously consistent with well-behaved virialized gas supported by thermal pressure in a hydrostatic halo.  In a complementary analysis of \ion{O}{6} absorption in 14 mostly sub-L$^*$ halos, \cite{mathes14} find a mass dependence for \ion{O}{6} kinematics. In their study, the velocities of \ion{O}{6} absorbers relative to the galaxy systemic velocities are much higher for galaxies with M$_{\rm halo}$ $<$ 10$^{11.5}$ M$_{\odot}$. There are several possibilities for such a large dispersion in velocities. For example, in \cite{oppenheimer16} the \ion{O}{6}-bearing gas traces extended $T \sim 10^{5.5}$ K gas mostly between $1-2$ R$_{\rm vir}$ around such halos, which would provide such a kinematic offset. 

Additionally, \cite{fielding16} present an idealized, three-dimensional hydrodynamic calculation in which they explore the intertwined roles of cosmological gas accretion and large-scale, cooling galactic winds for three fiducial halos at 10$^{11}$, 10$^{11.5}$, and 10$^{12}$ M$_{\odot}$. The gas properties in the lower mass halos are considerably more effected by the feedback from the galaxy.  Below their critical mass (10$^{11.5}$ M$_{\odot}$; offset from ours by 0.5 dex), \cite{fielding16} note that the cooling time of virialized gas is relatively short, and the halo gas is largely supported by turbulence and ram pressure generated from the vigorous feedback instead of the more typical thermal pressure thought to support virialized halos with M$_{\rm halo}$ $>$ 10$^{12}$ M$_{\odot}$. The turbulence would impact the line widths, and the large-scale motions would induce a sloshing of gas in the halo that could produce large velocity offsets from the galaxy systemic velocity. Such a picture may help to explain why the $|\Delta$v$|$ $>$ 100 km s$^{-1}$ for the `no-low' \ion{O}{6} absorbers in the lower mass halos. Future studies by both observers and theorists should explore the kinematics of virialized gas in halos with masses $<$ 10$^{12}$ M$_{\odot}$. 

The mean $\novi$ for the `no-low' absorbers is 10$^{14.2}$ cm$^{-2}$ (compared to 10$^{14.5}$ cm$^{-2}$ for the `broad' \ion{O}{6} absorbers). Their ratios $\nnv/\novi$ indicate that \ion{O}{6} represents 15\% of all the oxygen in this gas (Figure \ref{fig:collisional}), implying a total column, N$_{\rm O}$ $=$ 10$^{15}$ cm$^{-2}$. We follow the simple formulation by \cite[e.g][]{tumlinson11, werk14} of turning a column density into a mass surface density in order to find the total mass of the `virialized' halo gas. We allow the gas to have a metallicity between solar and a tenth solar \citep{asplund09}.  Assuming this gas fills the 150 kpc radius of the low-mass subset of COS-Halos galaxies, we find M$_{\rm gas}$ $\approx$ 10$^{9 - 10}$ M$_{\odot}$, where the higher value corresponds to the lower metallicity. For a halo mass of 10$^{11.5}$ M$_{\odot}$ that has its full cosmological share of baryons ($\sim$17\%), this mass represents anywhere between 5 and 50\% of the galaxy halo baryons. We note that in the \cite{fielding16} model, this virial temperature gas is cooling rapidly, which results in a complete loss of thermal pressure support, which leads to dramatic winds and subsequent cooling shocks at T $\approx$ T$_{\rm vir}$. In contrast, the model presented by \cite{oppenheimer16} can account for the `no-low' absorbers with a low-density, $\sim$10$^{5.5}$ K substrate mainly at $1-2$ R$_{\rm vir}$ that has cooling times comparable to the Hubble time.

\subsection{Photoionization by Local Sources}

The measured upper limits on the ratio $\nnv/\novi$ in the highly ionized gas strongly disfavor origin of \ion{O}{6} in gas photoionized by the EUVB only for all kinematic subtypes. To emphasize this point, we constructed a toy model including a black-body radiation component from local sources and found that luminosities with values $\nu$L$_{\nu}$ $>$ 10$^{40}$ erg s$^{-1}$ at $\sim$10 Ryd should be produced by our galaxies in order to match the data. This value corresponds to radiation exceeding the level supplied by the EUVB by a factor of 100 at energies above 10 Rydberg for a gas density of log n$_{\rm H}$ $=$ $-3.5$ cm$^{-3}$. In this phenomenology, \ion{O}{6} is under-abundant around non star-forming galaxies because there are not enough high energy photons escaping from the disk of the galaxy to ionize oxygen into \ion{O}{6}.  A major implication of this model is that the dramatic release of these high energy photons from the disk  photoionizes the primary coolants of the halo gas, thus increasing the cooling time of the gas by up to two orders of magnitude \citep{cantalupo10}.  In other words, photoionization from local sources in the galaxy (distinct from SN feedback) would then strongly regulate the rate at which galaxies acquire their gas. This model additionally requires a vast revision of the standard model where quasars dominate the EUVB at $\sim$10 Ryd. 

We explored which sources of radiation may be viable candidates and found that both the hot ISM heated by supernovae and supersoft X-ray sources (SSSs) may produce sufficient high energy radiation at large R, $\sim$2 orders of magnitude larger than the EUVB at high energies. Future studies will address how to better constrain the shape and intensity, and thus the physical origin of the required radiation field. There are two benefits of this model: smaller OVI path lengths and SFR/R$^2$ correlation is explained. However, several challenges face the model. For one, the implied extra radiation has dramatic implications on the EUVB itself. Geometric considerations and the assumption that the ionizing emissions from the associated galaxy trace star formation or stellar mass prohibits more than an order unity enhancement at $>$ 100 kpc in proximity radiation over the background at all wavelengths \citep[][Upton Sanderbeck et al. in prep]{miraldaescude05}.  Furthermore, this model cannot easily explain the diversity of low and high ion correspondence, seen in both the `narrow' and `broad' \ion{O}{6}.



\subsection{Shocks and Radiatively Cooling Gas Behind Fast Winds}

  Galaxy-scale winds are a natural model to address for the CGM, especially given their observed ubiquity in star-forming galaxies \citep[e.g.][]{martin12, rubin14} and their potential to scale with SFR/R$^2$ \citep{borthakur13}. Recently, \cite{thompson16} presented a detailed picture in which initially hot, adiabatic outflows radiatively cool on large scales ($\sim$ 100 kpc) and timescales less than a Hubble time. These winds directly give rise to cool, photoionized gas clouds that precipitate out \citep[see also:][]{wang95, efstathiou00}. This analytical model solves several problems relating to the CGM gas. In particular, it does not require that the cool gas be in pressure equilibrium with a virialized hot halo, consistent with observations that seem to rule this pressure-equilibrium out \cite{werk14}. Because the cool $\sim$10$^{4}$ K clouds arise from the radiatively cooling gas bearing \ion{O}{6}, we would expect their velocities to be similar. This physical scenario would likely give rise to a range of observed velocities and line-widths, depending on impact parameter, evolutionary stage of the wind, and the angle of viewing the shocked and post-shocked material. We consider this model promising, but must await its predictions on the kinematic signatures one might observe in low and high ionization state gas before coming to a conclusion on its consistency with observations. 

\section{Conclusions}

Conclusions regarding the origin and fate of circumgalactic gas are inextricably linked to the initial assumptions we make about the physical processes that determine its ionization state. In this work we have demonstrated that the kinematics of the highly-ionized gas in addition to gas column density ratios,  contain a wealth of information useful for constraining physical models of the CGM. We have examined a number of equilibrium and non-equilibrium ionization models that predict gas characterized by strong \ion{O}{6} absorption. The constraints from the COS-Halos absorption-line column density and kinematics measurements strongly disfavor many of the models considered.  At least some fraction of \ion{O}{6} appears to represent halo gas at the virial temperature, while most of the total column of \ion{O}{6} may result from either gas photoionized primarily by local high-energy sources or gas radiatively cooling on large scales behind a multiphase, fast wind. The latter two models imply very different ionization states and physical origins for the gas. Each model has its own set of strengths and weaknesses. 

Successful models of the CGM must account for:  (1) the velocity correspondence between the low-ionization state, photoionized gas and the \ion{O}{6} absorption; (2)  \ion{O}{6} column densities $\gtrsim$ 10$^{14}$ cm$^{-2}$ and highly variable line widths, 10 km s$^{-1}$ $<$ $b$ $<$ 100 km s$^{-1}$; and (3) the absence of \ion{O}{6} around non-star-forming galaxies and related tight correspondence between $\novi$ and SFR/R$^2$. For photoionization models to progress, future studies must focus on a detailed treatment of both the origin and long-term survival of these richly structured clouds and the sources of ionizing radiation.  A different issue faces the analytic, phenomenological, and hydrodynamical  models that generate significant \ion{O}{6} by collisional ionization. These models often provide a highly detailed physical treatment of the gas dynamics, origin, and ionization state, but make limited predictions with respect to the full range of possible observational measurements.  Future insight on the origins of \ion{O}{6} will come from comparing detailed observational analyses of gas kinematics with the model predictions for  the line-of-sight kinematics.

\section{Acknowledgements}
   
  Support for this work was provided by NASA through program GO11598,
  and through Hubble Fellowship grant \# 51332 from the Space
  Telescope Science Institute, which is operated by the Association of
  Universities for Research in Astronomy, Inc., under NASA contract
  NAS 5-26555.  B. Oppenheimer's contributions to this work are supported through the NASA ATP14-0142 grant.  SC gratefully acknowledges support from Swiss National Science Foundation grant PP00P2\_163824. JKW acknowledges the Esalen Institute for providing a stunningly beautiful setting for writing a large portion of the text. The authors sincerely thank Blair Savage who provided thorough and insightful referee reports that significantly improved the paper. Specifically, he encouraged the detailed kinematic analysis, which was largely absent from the initially-submitted version of this paper. JKW would like to especially thank Jonathan Stern, Rongmon Bordoloi, Renyue Cen, Charles Danforth, Joe Hennawi, Sowgat Muzahid, Brian O'Shea, Paul Shapiro, Todd Thompson, Julianne Dalcanton, Sarah Tuttle, and Evgenii Vasiliev, for very useful exchanges regarding ionization processes, pressure-support in galaxy halos, and \ion{O}{6} in general. Additionally, JKW would like to acknowledge UW graduate student Hannah Bish for her helpful suggestion of the term `no-low.' 

{{\it Facilities:} \facility{HST: COS} \facility{Keck: LRIS} \facility{Magellan: Mage} }

\bibliographystyle{apj}
\bibliography{cgmmass_all,extra_refs}


\end{document}